\documentclass[aps,prl,twocolumn,superscriptaddress,tightenlines]{revtex4-2}
\usepackage{graphicx}
\usepackage{dcolumn}
\usepackage{bm}
\usepackage{epsfig}
\usepackage{amsmath}
\usepackage{amsfonts}
\usepackage{amssymb}
\usepackage{empheq}
\usepackage{color}
\usepackage{bbm}
\usepackage[caption=false]{subfig}
\usepackage{placeins}
\usepackage{setspace}
\usepackage{mathtools}
\usepackage{extarrows}
\usepackage{soul}
\usepackage{physics}
\usepackage{mathrsfs}
\usepackage{changes} 
\definechangesauthor[name={Per cusse}, color=red]{per} 
\usepackage[colorlinks,citecolor=blue,linkcolor=blue]{hyperref}

\usepackage{orcidlink}
\newcommand{\customsqrt}[1]{\sqrt{\smash[b]{#1}}}

\definecolor{mybo}{cmyk}{0,0,0.10,0}

\begin{document}
	\title{Chiral Interaction Induced Near-Perfect Photon Blockade}
	\author{Zhi-Guang Lu~\orcidlink{0009-0007-4729-691X}}
	\affiliation{School of Physics and Institute for Quantum Science and Engineering, Huazhong University of Science and Technology and Wuhan institute of quantum technology, Wuhan, 430074, China}
	\author{Ying Wu}
	\affiliation{School of Physics and Institute for Quantum Science and Engineering, Huazhong University of Science and Technology and Wuhan institute of quantum technology, Wuhan, 430074, China}
	\author{Xin-You L\"{u}}\email{xinyoulu@hust.edu.cn}
	\affiliation{School of Physics and Institute for Quantum Science and Engineering, Huazhong University of Science and Technology and Wuhan institute of quantum technology, Wuhan, 430074, China}
	
	\date{\today}
	\begin{abstract}
		Based on the scattering matrix method, we theoretically demonstrate that the chiral interaction can induce the almost perfect photon blockade (PB) in the waveguide-cavity quantum electrodynamics (QED) system. The mechanism relies on the multi-photon paths interference within the waveguide, which is clearly shown by the analytic parameter regime for $g^{(2)}(0)\approx0$. When $N$ cavities are introduced into the system, there are $N$ optimal parameter points accordingly for the almost perfect PB, where the required lowest chirality decreases exponentially with increasing $N$, and these optimal points are robust against disorder in the system's frequencies. Under resonant driving and fixed chirality conditions, the output light depends solely on the parity of $N$ ($N\ge2$), with a coherent state emerging for even numbers of cavities and a single-photon state for odd numbers. Our work offers an alternative route for achieving almost perfect PB effects with high single-photon transmission by employing the chirality of system, with potential application in the on-chip single-photon source with integrability.
	\end{abstract}
	\maketitle
	
	Photon blockade (PB) \cite{PhysRevLett.79.1467}, i.e., the blockade of subsequent photons after absorbing the first one \cite{PhysRevA.46.R6801,PhysRevA.49.R20,1996Quantum}, has attracted significant theoretical and experimental attention in recent years, due to its fundamental interest in exploring the quantum behavior of light \cite{PhysRevLett.70.2269,PhysRevLett.70.2273, PhysRevA.65.063804} and its potential applications in quantum information processing \cite{Buluta_2011,bennett_quantum_2000,Lodahl_2018}. Generally, PB can be categorized into conventional \cite{PhysRevLett.79.1467} and unconventional types \cite{PhysRevLett.104.183601,PhysRevA.83.021802}. Conventional PB effect is induced by the anharmonicity in the spectrum of physical systems, requiring strong optical nonlinearities. This regime can be achieved in various systems such as cavity quantum electrodynamics (QED) \cite{PhysRevLett.103.150503, faraon_coherent_2008,birnbaum_photon_2005, PhysRevLett.127.240402, reinhard_strongly_2012, PhysRevA.69.035804}, superconducting circuits \cite{PhysRevLett.106.243601,PhysRevLett.107.053602,PhysRevA.94.053858}, optomechanical resonators \cite{PhysRevLett.107.063601,PhysRevLett.107.063602,PhysRevA.94.063853,kim_low-dielectric-constant_2016}, and others \cite{PhysRevLett.108.183601, PhysRevB.87.235319}. In contrast, unconventional PB effect is caused by the interference effect among different transition paths within physical systems and can be achieved in weakly nonlinear regimes. 
	
	With the rapid development of quantum computing, quantum simulation, and other emerging quantum technologies \cite{RevModPhys.79.135,PhysRevLett.108.206809,Noh_2017,knill_scheme_2001,RevModPhys.83.33}, the near-perfect PB effect, i.e., $g^{(2)}(0)\approx0$, becomes increasingly significant for a variety of precision quantum devices, such as high-precision quantum gates \cite{PhysRevApplied.13.044013,tiarks_photonphoton_2019,hacker_photonphoton_2016}, high-efficiency single-photon devices \cite{santori_indistinguishable_2002,09500340.2012.744192,PhysRevLett.118.190501}, and other quantum devices \cite{PhysRevX.5.021025,PhysRevApplied.7.024028,PhysRevApplied.7.034031,PhysRevX.7.031001,PhysRevLett.120.023601,PhysRevLett.121.153601}. Waveguide QED systems offer an important platform for achieving these precision quantum devices due to their unique design and operating principles, including easy integration \cite{sciadv.abc8268, wang_integrated_2020}, tunability, and high efficiency \cite{bock2016highly}. In particular, it is possible to experimentally achieve chiral interactions \cite{lodahl_chiral_2017, Tan:22, sollner_deterministic_2015} between a waveguide and cavity (or atom), where the emission or scattering of photons from the cavity (or atom) strongly depends on the photon's propagation direction within the one-dimensional waveguide. Here the directionality originates from the interaction of elliptical dipoles with finely structured light fields of nanophotonic systems \cite{PhysRevResearch.4.023082}, such as nanowaveguides \cite{science.1257671,PhysRevB.95.121401}, resonators \cite{Schneeweiss:17, acsphotonics.8b01555}, and photonic-crystal (PhC) waveguides \cite{PhysRevLett.115.153901,le_feber_nanophotonic_2015,sollner_deterministic_2015}. It has been demonstrated that chiral interactions can induce significant quantum effects, such as the directionality of spontaneous emission \cite{sollner_deterministic_2015, coles_chirality_2016} and the chiral Purcell effect \cite{PhysRevLett.114.203003}. A natural question arises: could these interactions significantly influence the statistical properties of outgoing light, ultimately leading to the near-perfect PB effect?
	
	Here we propose how to utilize chiral interactions to achieve an almost perfect PB effect in a coupled waveguide-cavity QED system, which is experimentally feasible \cite{acsphotonics}. Using the scattering theory, we provide a compact analytical expression for the $n$th-order correlation functions with zero-time delay and demonstrate the destructive interference of multi-photon paths induced by chiral interactions. This interference effect can induce near-perfect single-photon blockade (1PB), two-photon blockade (2PB), and photon-induced tunneling (PIT) by modulating the chiral waveguide-cavity interaction. For our system including $N$ cavities, we find that there are $N$ optimal parameter points for the near-perfect PB associating with the high single-photon transmission due to the increased scattering paths. The lowest chirality required for near-perfect PB decreases exponentially as the number of cavities increases and is robust against disorder in the system's frequencies. We also find that under resonance driving conditions, the near-perfect PB appears in systems with an odd number of cavities and disappears in systems with an even number of cavities. Note that, our main results can not be obtained in the previous works\,\cite{PhysRevLett.121.143601, prasad_correlating_2020}, exploring the systems with many emitters directly coupled to the waveguide, since the different physical mechanisms and models are introduced into our work\,\cite{supp}. Our work opens the door to exploring the crossover between chiral quantum optics and quantum state engineering, offering potential applications in designing new types of high-purity single-photon devices with high brightness. \nocite{PhysRevLett.121.043601, PhysRevLett.121.043602, PhysRevA.12.1919, Pathria1996, lidar_lecture_2020, johansson_qutip_2012, johansson_qutip_2013, manzoni_simulating_2017, qt}
	
\begin{figure}
		\includegraphics[width=8.5cm]{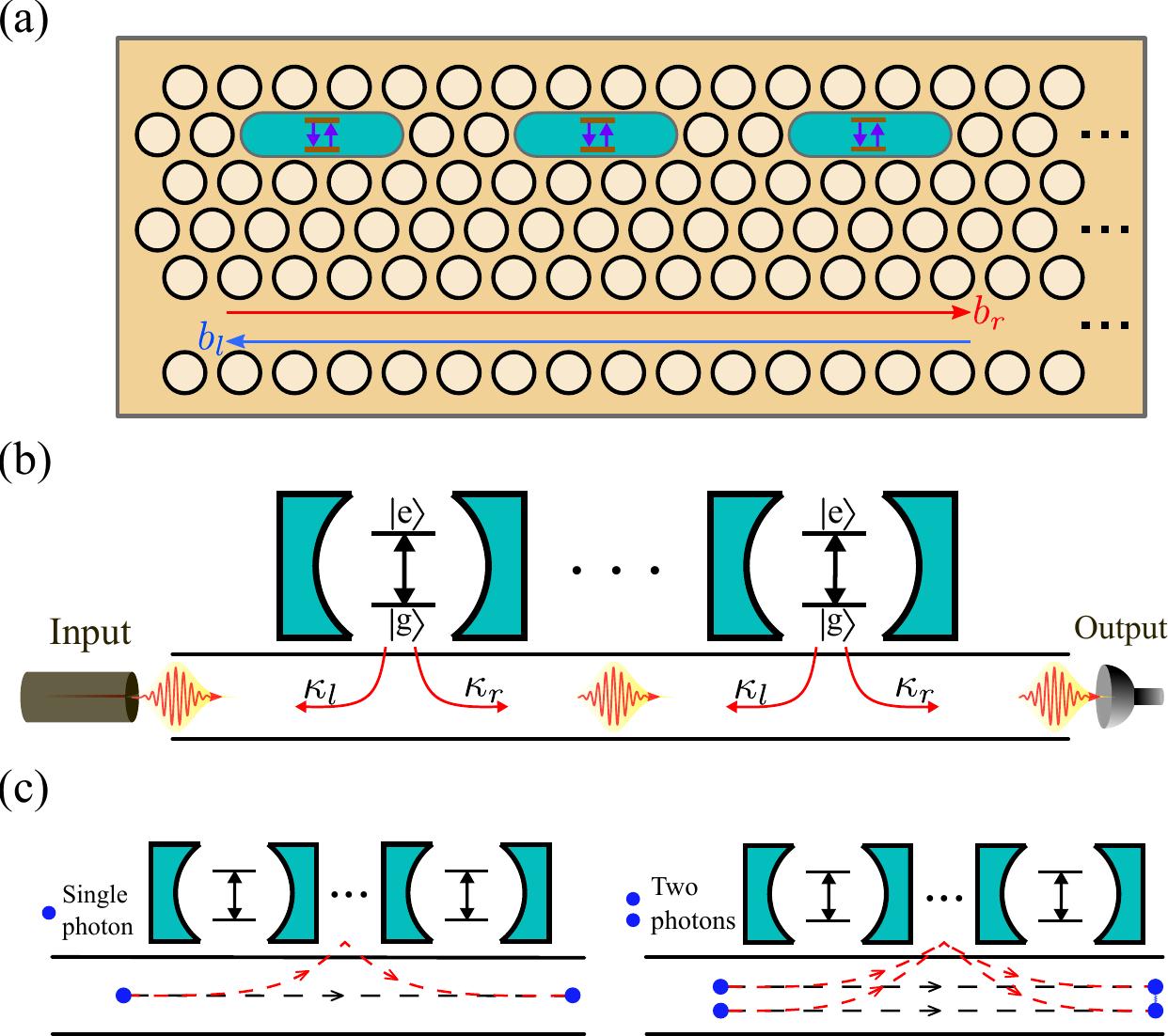}\\
		\caption{(a) The implementation of a coupled waveguide-cavity QED system with a PhC structure, including multiple point-defect H1-type PhC cavities coupled to semiconductor quantum dots, and a line-defect waveguide. (b) The general model of a coupled waveguide-cavity QED system. Each cavity can emit photons to the left and right propagating waveguide modes. The chirality of the cavity-waveguide interaction refers to the asymmetry in the corresponding decay rates, i.e., $\kappa_l\neq\kappa_r$. (c) Illustrations of the single- and two-photon scattering processes within the waveguide.}\label{fig-m1}
	\end{figure}

	\emph{Model and Hamiltonian}.---As shown in Fig.\,\ref{fig-m1}(a), the system consists of multiple PhC cavities, each denoted by the annihilation operator $a_j$, and each cavity is coupled to a two-level atom, denoted by the lowering operator $\sigma_j\equiv|g\rangle_j\langle e|$, with the coupling strength $\text{g}$. The cavities are side-coupled to a PhC waveguide, and in this configuration, the scattering paths of photons are markedly different from those in the case of a direct-coupled waveguide \cite{supp}. The waveguide mode is described by a frequency-dependent annihilation operator $b_\mu(\omega)$, where $\mu=l$ ($r$) represents the left (right) propagating mode. This operator satisfies the commutation relation $[b^{}_\mu(\omega),b_{\mu^\prime}^\dagger(\omega^\prime)]=\delta_{\mu,\mu^\prime}\delta(\omega-\omega^\prime)$. Then, the total Hamiltonian of the coupled cavity-waveguide system is given by $H_{\text{tot}}=H_{\text{sys}}+H_{\text{wg}}+H_{\text{int}}$ with ($\hbar=1$) \cite{PhysRevA.91.042116}
	\begin{align}
		H_{\text{sys}}&=\sum_{j=1}^{N}[\omega_ca_j^\dagger a_j^{}+\omega_e\sigma_j^\dagger\sigma_j^{}+\text{g}(a_j^\dagger\sigma_j^{}+\sigma_j^\dagger a_j^{})]\nonumber,\\
		H_{\text{wg}}&=\sum_{\mu=l,r}{\int\dd\omega\ \omega b^\dagger_\mu(\omega)b^{}_\mu(\omega)},\label{eq-m2}\\
		H_{\text{int}}&=\sum_{j=1}^N\sum_{\mu=l,r}\sqrt{\frac{\kappa_\mu}{2\pi}}\int\dd\omega[b_\mu^\dagger(\omega)a^{}_je^{-i\omega x_j/v_\mu}+\text{h.c.} ],\nonumber
	\end{align}
	where $\omega_c$ is the cavity resonance frequency, $\omega_e$ is the atomic transition frequency, $\kappa_l$ and $\kappa_r$ are the decay rates into the left ($v_l<0$) and right ($v_r>0$) propagating waveguide modes, respectively. In a realistic platform, the cavity and atom also have intrinsic dissipation that decays into non-waveguide loss channels, and the impact of the intrinsic dissipation is discussed in the Supplemental Materials (SM) \cite{supp}. Here, $v_\mu$ denotes the group velocities in the waveguide (assuming $\abs{v_r}=\abs{v_l}=v$), and $x_j$ denotes the position of the $j$-th cavity. The chirality is defined as $\alpha\equiv\kappa_l/\kappa_r$ ($\alpha\in[0,1]$), with the perfectly chiral and nonchiral cases corresponding to $\alpha=0$ and $\alpha=1$, respectively. In the following, we take the spacing between neighboring cavities to be constant $d$ in the waveguide, and the atomic transition frequency to resonate with the cavity mode frequency, i.e., $\omega_c=\omega_e$. 
	
	Based on the scattering theory, we regard the cavity QED system as a potential field (or scatterer) \cite{PhysRevB.98.144112} influencing the propagation of incident photons within the waveguide. The statistical properties of the output state can be completely described by the scattering processes of photons. As shown in Fig.\,\ref{fig-m1}(c), upon the incidence of a single photon along the waveguide, there are two possible paths for it to be scattered into the right-side output port. One path is free propagation along the waveguide, shown by the black dotted line. The other path involves the photon being first absorbed by the cavity and then emitted into the right-side output port (i.e., right-moving mode), shown by the red dotted line. Similarly, for the incidence of two photons without interaction, there are four possible paths: two photons can be scattered into the right-side output port, and photon-photon interactions can be induced by the presence of the potential field, as shown in the two red dotted lines. Meanwhile, the two black dotted lines represent the free propagating path without the influence of the potential field, which does not induce photon-photon interactions. The scattering matrix (S-Matrix) method \cite{SM1,SM2,SM3,SM4,SM5,SM6,SM7,SM8,SM9,SM10,SM11,SM12,SM13} is optimal for describing the scattering processes of photons, and we only need to discuss few-photon scattering processes for our purpose. For simplicity, we consider the incident single-photon and two-photon states with frequencies $\omega_d$. The output states are $|\Psi_{\text{out}}^{(1)}\rangle^{r}_{\omega_d}=\int\dd p S^{ rr}_{p;\omega_d}b^\dagger_r(p)\ket{0}$ and $|\Psi_{\text{out}}^{(2)}\rangle^{r}_{\omega_d}=\int\frac{\dd p_1\dd p_2}{2!\sqrt{2!}} S^{ rr}_{p_1p_2;\omega_d\omega_d}b^\dagger_r(p_1)b^\dagger_r(p_2)\ket{0}$ \cite{supp}, where $S_{p;\omega_d}^{rr}, S_{p_1p_2;\omega_d\omega_d}^{rr}$ are the single- and two-photon S-Matrix elements, respectively. The S-Matrix elements can be completely characterized by the effective non-Hermitian Hamiltonian \cite{supp}
	\begin{align}
		H_{\text{eff}}\!=\!H_{\text{sys}}&\!-\!i\sum_{j=1}^N\frac{\kappa}{2}a_{j}^\dagger a_{j}^{}\!-\!i\sum_{j>k}^N(\kappa_ra_{j}^\dagger a_{k}^{}+\kappa_la_{k}^\dagger a_{j}^{}),\label{eq-m4}
	\end{align}
	where $\kappa=\kappa_l+\kappa_r$. Note that the phase $\phi=\omega_cd/v$ takes the value $2m\pi$ with $m\in\mathbb{N}$, corresponding the case of mirror configuration. Other cases are analyzed in the SM \cite{supp}.
	
	\emph{Analytical nth-order correlation functions}.---We consider a weak coherent state with amplitude $\eta$ entering into the waveguide from the left-side port, as shown in Fig.\,\ref{fig-m1}(b), and thus the initial state is given by $|\psi_{\text{in}}\rangle=\mathcal{N}\sum_{n=0}^\infty\frac{\eta^n}{\sqrt{n!}}|\Psi^{(n)}_{\text{in}}\rangle^{r}_{\omega_d}|0\rangle_{l}|0\rangle_s$, where $|\Psi^{(n)}_{\text{in}}\rangle^{r}_{\omega_d}=[b_r^{\dagger n}(\omega_d)/\customsqrt{n!}]|0\rangle_{r}$ denotes the Fock state of $n$ photons with frequency $\omega_d$ in the right-moving waveguide mode. Here, $|0\rangle_{l(r)}$ is the photonic vacuum state in the left- (right-) moving waveguide mode, $|0\rangle_s$ is zero-excitation eigenstate of the effective Hamiltonian, and $\mathcal{N}$ is a normalization factor. According to scattering theory, the output state satisfies $|\psi_{\text{out}}\rangle=S|\psi_{\text{in}}\rangle$, where $S$ is the scattering operator. In the weak drive limit, the $n$th-order equal-time correlation functions and the single-photon transmission in the output channel are given by \cite{lu2023analytical}
	\begin{align}
		g_{rr}^{(n)}(0)
		&=\lim\limits_{\abs{\eta}\to0}\frac{\langle\psi_{\text{out}}|b_r^{\dagger n}(t)  b_r^n(t)|\psi_{\text{out}}\rangle}{\ \langle\psi_{\text{out}}| b^\dagger_r(t)b^{}_r(t)|\psi_{\text{out}}\rangle ^n}
		=\frac{|\mathbb{P}_n|^2}{\ \ |\mathbb{P}_1|^{2n}},\label{eq-m6}\\
		T&=\lim\limits_{\abs{\eta}\to0}\frac{\langle\psi_{\text{out}}|b_r^{\dagger }(t)  b_r(t)|\psi_{\text{out}}\rangle}{\ \langle\psi_{\text{in}}| b^\dagger_r(t)b^{}_r(t)|\psi_{\text{in}}\rangle}\label{mT}
		=2\pi|\mathbb{P}_1|^2,
	\end{align}
	where $\mathbb{P}_n$ denotes the $n$-photon total probability amplitude, representing the amplitude of probing $n$-photon with zero-time delay in the right-side output port. Considering all the possible scattering paths of $n$ incident photons, the total amplitudes are actually a linear combination of the amplitude of each possible path, i.e.,
	\begin{eqnarray}
		\begin{aligned}\label{eq-m7}
			\mathbb{P}_n&=\frac{1}{\customsqrt{(2\pi)^n}}\sum_{m=0}^n{C_n^mP^{N}_m\mathcal{I}^{}_{n-m}}.
		\end{aligned}
	\end{eqnarray}	
	
	\noindent Here $P_{n}^N$ ($P^N_0=1$) denotes as the $n$-photon probability amplitude, representing the amplitude that $n$ photons are first absorbed by the system including $N$ cavities and then emitted into the right-moving waveguide mode, and $C_n^m$ is the binomial coefficient, indicating all possible paths. Additionally, $\mathcal{I}_{n}=\exp(-in\omega_d t)$ denotes the probability amplitude of $n$ photons freely propagating along the waveguide, and the incoming $n$ photons have $2^n$ possible paths if we number each photon, i.e., $\sum_{m=0}^nC_{n}^m=2^n$, where the number $2$ represents the two kinds of scattering paths for each single photon, as shown in Fig.\,\ref{fig-m1}(c). Physically, since the scattering path with a probability amplitude of $P_{n}^N$ is governed by the transition paths, the superposition of these scattering and transition pathways gives rise to the unique interference effect in the waveguide QED system, potentially inducing the near-perfect PB. Significantly, for our system, the single-photon probability amplitude has an analytic expression given by \cite{supp}
	\begin{align}\label{eq-m8}
		P_{1}^N=-\frac{1-\exp(iN\theta)}{1-\alpha \exp(iN\theta)}\ \mathcal{I}_1,
	\end{align}
	where $\tan(\theta/2)=\Delta(1-\alpha)\kappa/[2(1+\alpha)(\Delta^2-\mathrm{g}^2)]$, with $\Delta=\omega_c-\omega_d$ denoting the detuning between the cavity and the incident photon frequencies. 

	\begin{figure}
		\includegraphics[width=8.5cm]{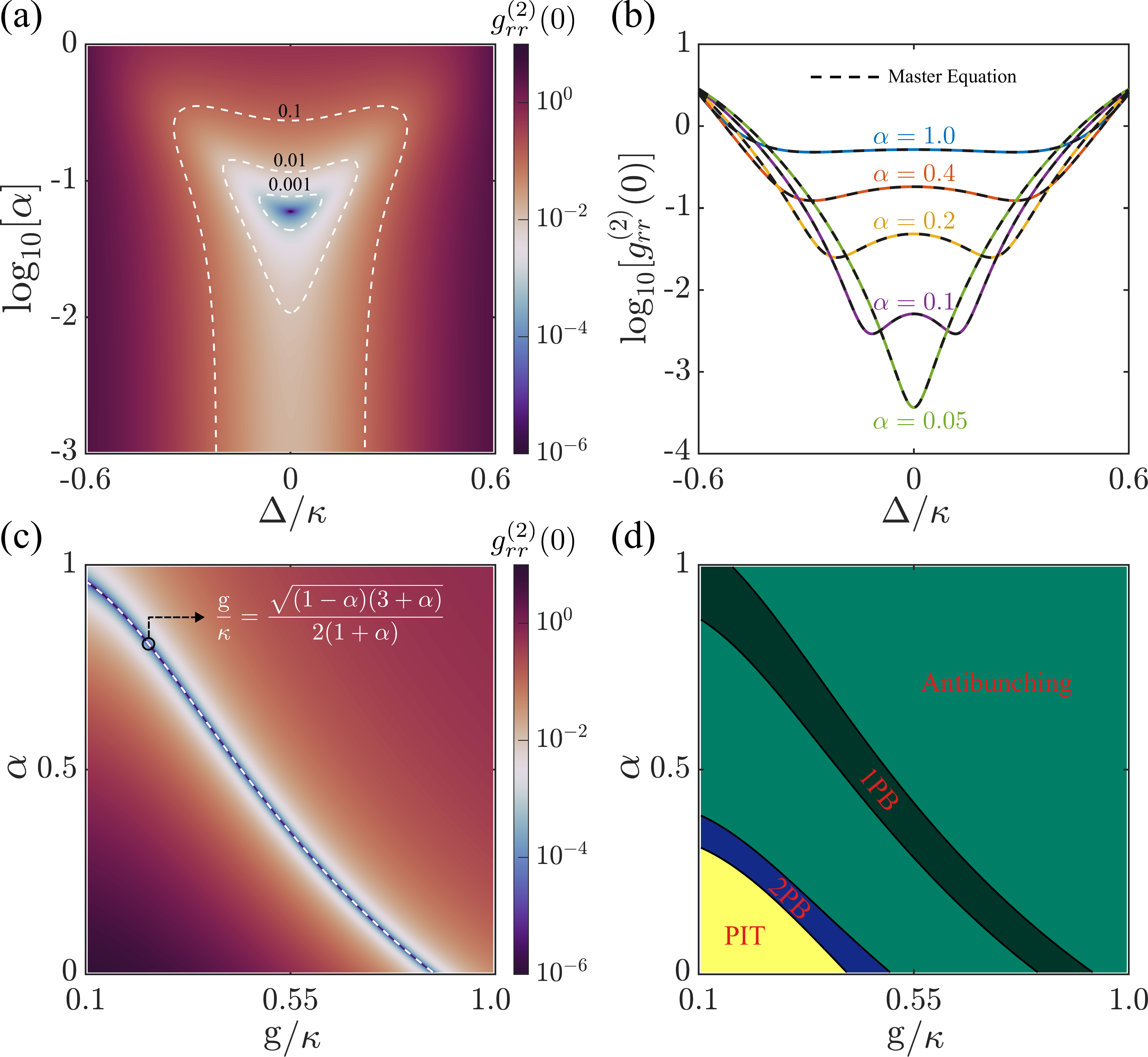}\\
		\caption{The equal-time second-order quantum correlation $g_{rr}^{(2)}(0)$ as a function of the chirality $\alpha$ and (a) the detuning $\Delta/\kappa$, (c) the coupling strength $\text{g}/\kappa$. The dashed lines in (a) represent the contour lines for $g_{rr}^{(2)}(0)=0.1,\,0.01,\,0.001$, and the dashed line in (c) denotes the parameter curve corresponding to $g_{rr}^{(2)}(0)=0$. (b) $g_{rr}^{(2)}(0)$ versus the detuning $\Delta/\kappa$ for different chirality. The dashed lines are the numerical results obtained from the master equation with a driving strength $\eta\sqrt{\kappa_r/2\pi}=10^{-3}\kappa$, while the solid lines show the analytical results from the scattering matrix. (d) Phase diagram illustrating the statistical properties of output light. The system parameters are $\text{g}/\kappa=0.8$ in (a, b) and $\Delta=0$ in (c, d).}\label{fig-m2}
	\end{figure}	
    
	\emph{Near-perfect PB and phase diagram}.---We first consider the effect of chirality on PB in a single-cavity case ($N=1$). To obtain the second- and third-order quantum correlations, we need to compute $P^{1}_n$ for $n=1,2,3$ according to Eq.~(\ref{eq-m6}) and Eq.~(\ref{eq-m7}). The two- and three-photon probability amplitudes are given by
	\begin{align}
		P^{1}_2&=\frac{\text{g}^2+\Delta\Delta_c+\Delta^2}{\prod_{n=0}^{1}[(n\Delta_c^2+\Delta\Delta_c-\text{g}^2)(1+\alpha)/\kappa]}(i\mathcal{I}_1)^2,\label{eq-m10}\\
		P^{1}_3&=\frac{\prod_{n=0}^{2}(n\Delta_c+\Delta)+(4\Delta_c+3\Delta)\text{g}^2}{\prod_{n=0}^{2}[(n\Delta_c^2+\Delta\Delta_c-\text{g}^2)(1+\alpha)/\kappa]}(i\mathcal{I}_1)^3,\label{eq-m11}
	\end{align}
	where $\Delta_c=\Delta-i\kappa/2$. Figures \ref{fig-m2}(a) and \ref{fig-m2}(b) show that the second-order quantum correlation approaches to zero around $\Delta=0$ as $\alpha$ decreases, i.e., $g_{rr}^{(2)}(0)|_{\alpha=1}\approx1$ and $g_{rr}^{(2)}(0)|_{\alpha=0.05}\approx10^{-4}$. We know that the perfect PB, i.e., $g_{rr}^{(2)}(0)=0$, indicates that the two-photon total probability amplitude has exact zeros, i.e., $\mathbb{P}_2=0$. Thereby, the second-order quantum correlation is zero only when destructive interference between the two-photon paths occurs. Under the resonant condition $\Delta=0$, the optimal parameter curve for perfect PB is
	\begin{align}\label{eq-m12}
		\text{g}/\kappa=\frac{\sqrt{(1-\alpha)(3+\alpha)}}{2(1+\alpha)}\quad (0\le\alpha<1)
	\end{align}
	which corresponds to the white dotted line in Fig.\,\ref{fig-m2}(c). Equation~(\ref{eq-m12}) clearly demonstrates that the perfect PB effect only occurs when the decay rates of cavity are asymmetric, i.e., $\alpha<1$, which induces destructive interference of the two-photon paths. We also study some interesting quantum phenomena besides 1PB, such as 2PB and PIT. For the criterion of 2PB, the correlation functions must fulfill $g_{rr}^{(3)}(0)<1$ and $g_{rr}^{(2)}(0)\ge1$ \cite{PhysRevLett.121.153601}. The criterion for PIT is $g_{rr}^{(n)}(0)>1$ for $n=2,3$ \cite{PhysRevLett.121.153601}. Combining these criteria, we present the phase diagram describing the statistical properties of output light in Fig.\,\ref{fig-m2}(d). Notice that the criterion for 1PB corresponds to $g_{rr}^{(2)}(0)<0.01$ here. It is clearly shown that 2PB and PIT only appear in the region with chirality. This phase diagram provides an intuitive understanding of the statistical properties of outgoing light in different parameter regions.
	
\begin{figure}
		\includegraphics[width=8.5cm]{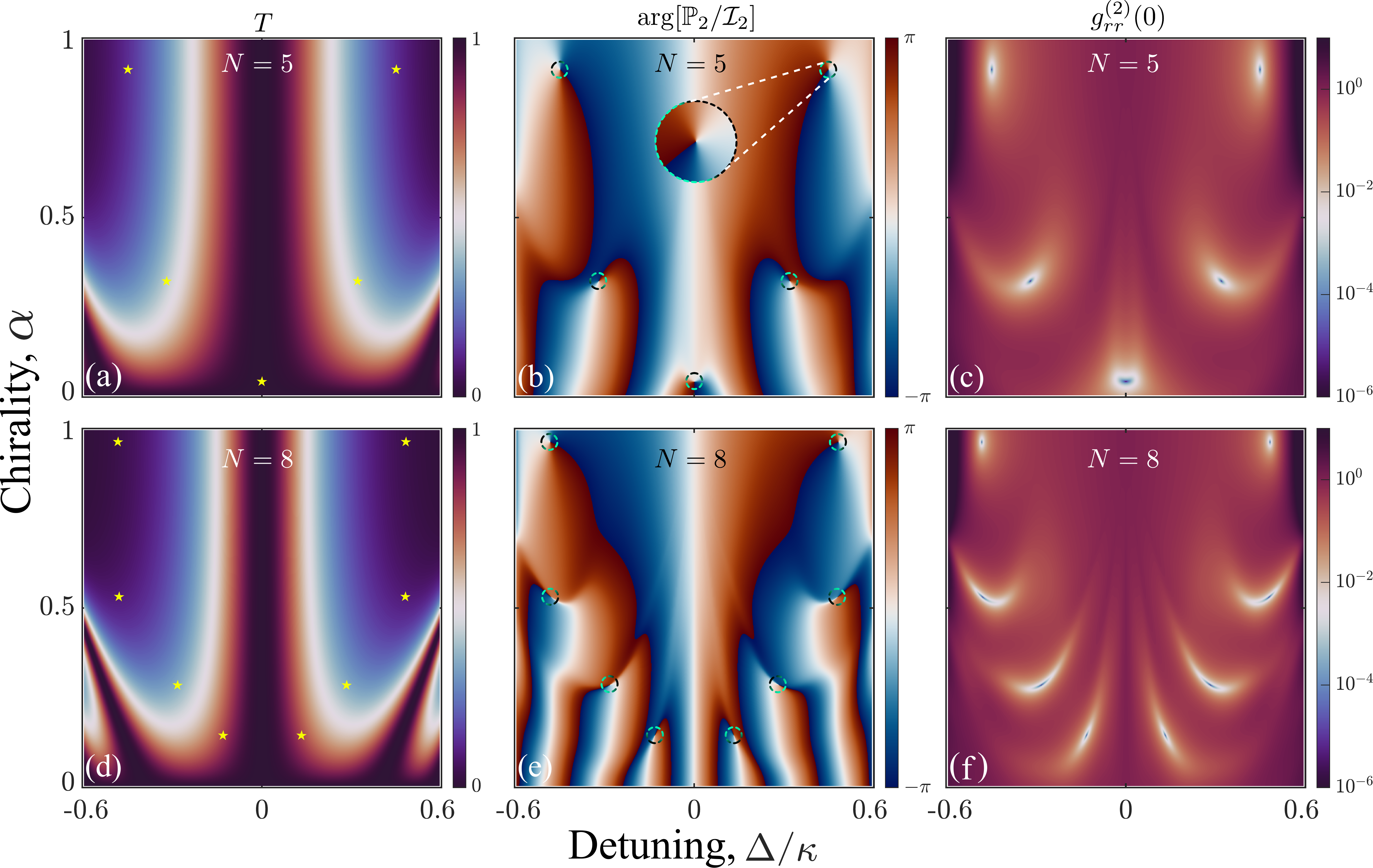}\\
		\caption{The figures show the single-photon transmission $T$ in (a, d), and (b, e) the complex argument of the time-independent two-photon total probability amplitude $\arg[\mathbb{P}_2/\mathcal{I}_2]$, and (c, f) the equal-time second-order quantum correlation $g_{rr}^{(2)}(0)$, versus the chirality $\alpha$ and the detuning $\Delta/\kappa$ for $N=5$ (top row) and $N=8$ (bottom row). The yellow stars in (a, d) and dashed circles in (b, e) correspond to the locations of near-perfect PB points in (c, f), respectively. The system parameters are the same as in Fig.\,\ref{fig-m2}(a).
		}\label{fig-m3}
	\end{figure}

	\emph{Multi-cavity system}.---As the number of cavities increases, the possible interference paths of the incoming photons among different cavities will increase accordingly, which may indirectly result in an increase in the number of perfect PB points. As shown by the analytic expression in Eq.~(\ref{eq-m8}), only when each cavity is chirally coupled to the waveguide can a determined single-photon probability amplitude potentially have $N$ roots in the parameters space $(\Delta, \alpha)$ due to the presence of $\exp(iN\theta)$. This is also true for the two-photon probability amplitude. To show this, Figs.\,\ref{fig-m3}(b) and \ref{fig-m3}(e) plot the complex argument of $\mathbb{P}_2/\mathcal{I}_2$ versus $\alpha$ and $\Delta$ for various $N$. We observe $N$ phase singularities (see the dashed circles) at discrete parameter points, corresponding to perfect PB points (analytic zeros). Therefore, Figs\,\ref{fig-m3}(b-f) confirm that the number of perfect PB points is consistent with the number cavities in the system, and a larger $N$ also leads to wider PB windows. Additionally, as shown in the stars of Figs.\,\ref{fig-m3}(a) and \ref{fig-m3}(d), we find that the single-photon transmission can also be high at the optimal detuning and chirality. Particularly, at zero detuning and with an odd number of cavities, the transmission is one. These findings highlight the progressive enhancement of PB effects with the expansion of the cavity-atom array. Our results also reveal another intriguing phenomenon: under the resonant condition, the perfect PB effect completely disappears for system with an even number of cavities ($N=2\ell$ for $\ell\in\mathbb{N}$), i.e., $g_{rr}^{(2)}(0)=1$. However, it remarkably reappears for system with an odd number of cavities ($N=2\ell+1$). To elucidate this phenomenon, we need to study the multi-photon scattering processes. First, we regard two adjacent-cavity systems as a unit cell for simplicity. By focusing on a system including a single unit cell and solving the $n$-photon S-Matrix elements, we find that $S^{rr}_{p_1\ldots p_n;\omega_d\ldots\omega_d}=n!\prod_{j=1}^{n}\delta(p_j-\omega_d)$. This means that the incoming $n$ photons freely propagate along the waveguide coupled to the system, resulting in $P_{n}^{2}=0$. Then, for a system with $\ell$ unit cells, the same result holds according to scattering theory, e.g., $P_n^{2\ell}=0$. Consequently, for a system with an even number of cavities, the output light still retain the statistical properties of the input state, i.e., $|\psi_{\text{out}}\rangle=|\psi_{\text{in}}\rangle$. However, for a system with an odd number of cavities, although the scattering processes of two photons align with the previously discussed behavior for the first $2\ell$ cavities, there exits multi-photon paths interference between the first $2\ell$ cavities and the last cavity, which leads to \cite{supp}
	\begin{align}\label{eq-m13}
		P_2^{2\ell+1}=\frac{4(1-\alpha)^2/(1+\alpha^2)}{4(\text{g}/\kappa)^2(1+\alpha)^2+(1-\alpha)^2}(i\mathcal{I}_1)^2.
	\end{align}
	Comparing with the system with an even number of cavities, Eq.~(\ref{eq-m13}) shows that near-perfect PB reemerges for an appropriate chirality. Note that, for the multi-cavity system, we have verified the correctness of S-Matrix method through matrix product states simulations \cite{supp}.

	\begin{figure}
		\includegraphics[width=8.5cm]{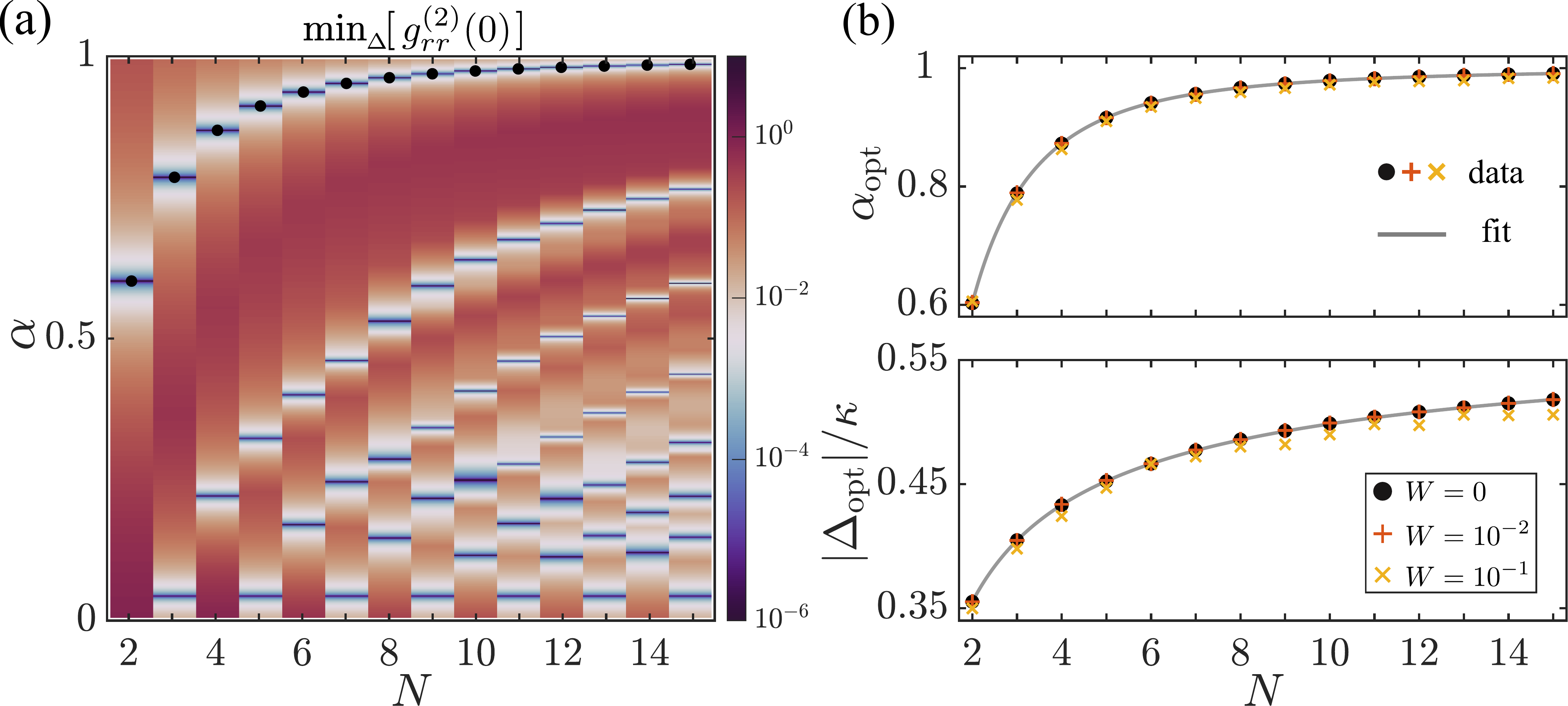}\\
		\caption{(a) The minimum second-order quantum correlation $\text{min}_{\Delta}[g_{rr}^{(2)}(0)]$ as a function of chirality $\alpha$ and the number of cavities $N$. Black dots indicate the required lowest chirality $\alpha_{\text{opt}}$. (b) Top: optimal chirality $\alpha_{\text{opt}}$ as a function of $N$. The gray line is a fit with the function $f(N)=1-N^{\eta}\sum_{i=1}^{2}c_i\exp(-a_iN)$ for $N>1$ (main: $a_1=0.456,a_2=0.089,c_1=1.103,c_2=0.294, \eta=-0.795$). Bottom: the corresponding optimal detuning $|\Delta_{\text{opt}}|/\kappa$ as a function of $N$. The gray line is a fit with the function $h(N)=0.593-c_3 N^{\gamma}$ for $N>1$ (main: $c_3=0.354, \gamma=-0.574$). Note in (b) that the dot, the plus sign, and the cross correspond to the average value computed with a total of $10^3$ instances of disorder with strength $W=0$, $W=0.01$, and $W=0.1$, respectively. The other parameters are the same as in Fig.\,\ref{fig-m2}(a).
		}\label{fig-m4}
	\end{figure}
	
	To extract more information regarding the perfect PB, we iterate over all the detuning for each $(\alpha, N)$ to obtain a minimum second-order quantum correlation. As shown in Fig.\,\ref{fig-m4}(a), each dark-blue transverse fringe corresponds to a perfect PB point, and the number of fringes equals $\lceil N/2\rceil$ for a given $N$. More importantly, we also study the effect of disorder in the system's frequencies, corresponding to the addition of random diagonal terms to the system's Hamiltonian, $H_{\text{sys}}\to H_{\text{sys}}+\sum_{j}(\epsilon_{c,j}a_j^\dagger a_j^{}+\epsilon_{e,j}\sigma_j^\dagger \sigma_j^{})$. We take $\epsilon_{\nu,j}/\kappa$ from a uniform distribution within the range $[-W, W]$ for each $j$th cavity ($\nu=c$) and emitter ($\nu=e$), where $W$ represents the disorder strength. For $W=0$, Fig.\,\ref{fig-m4}(b) shows that the lowest chirality required for achieving perfect PB exhibits an exponential decrease with $N$, i.e., $1-\alpha_{\text{opt}}\propto \exp(-a_iN) \text{ with } a_i>0$, and the corresponding detuning also obeys power law relationship, i.e., $0.593-\abs{\Delta_{\text{opt}}}/\kappa\propto N^\gamma \text{ with } \gamma<0$. Notably, even for $W\neq0$, as illustrated by the plus sign and the cross in Fig.\,\ref{fig-m4}(b), the optimal chirality and detuning are robust against disorder, showing minor shifts compared to a clean system. These intriguing results may stem from the exponential growth of interference paths, and the exponential term $\exp(iN\theta)$ in the single-photon probability amplitude~(\ref{eq-m8}) indirectly supports this idea. Consequently, as the interference paths increase exponentially, tiny shifts of these paths, induced by an exponentially weak chirality, can significantly impact the statistical properties of the output light, potentially leading to the emergence of perfect PB in large-$N$ system with extremely weak chirality. This property opens the door for significantly reducing the required chirality to obtain a high-purity single-photon source.
	
	\emph{Conclusion}.---In conclusion, we have demonstrated a condition for inducing near-perfect PB with high single-photon transmission when each cavity couples asymmetrically to the left- and right-propagating waveguide modes. Additionally, we have showed that the required lowest chirality for near-perfect PB decreases exponentially with the number of cavities. Given that our results are robust against disorder in the system's frequencies and exhibit strong immunity to the intrinsic cavity dissipation \cite{supp}, our work has the potential to realize scalable single-photon sources \cite{PhysRevLett.123.250503,sciadv.abc8268} using a mature PhC waveguide platform, which might significantly impact quantum information science applications. Furthermore. our findings may inspire further studies on other quantum effects induced by chiral coupling of cavities to waveguide, and the S-Matrix method we used may also become a promising tool for probing many-body (subradiant) states in (non-) Hermitian many-body systems.

    \emph{Acknowledgments}.---This work is supported by the National Science Fund for Distinguished Young Scholars of China (Grant No. 12425502), the National Key Research and Development Program of China (Grant No. 2021YFA1400700), and the Fundamental Research Funds for the Central Universities (Grant No. 2024BRA001). The computation was completed in the HPC Platform of Huazhong University of Science and Technology.

	%

\clearpage
\setcounter{secnumdepth}{2}
\onecolumngrid
\begin{center}
	{\Large \textbf{ Supplemental Material for\\
			``Chiral Interaction Induced Near-Perfect Photon Blockade"}}
\end{center}

\begin{center}
	Zhi-Guang Lu$^{1}$, Ying Wu$^{1}$, Xin-You L\"{u}$^{1,*}$
\end{center}
\begin{center}
\begin{minipage}[]{16cm}
    \small{\it
    \centering $^{1}$ School of Physics and Institute for Quantum Science and Engineering, Huazhong University of Science and Technology and Wuhan Institute of Quantum Technology, Wuhan 430074, China\\}
\end{minipage}
\end{center}
\setcounter{equation}{0}
\setcounter{figure}{0}
\setcounter{table}{0}
\setcounter{section}{0}
\makeatletter
\renewcommand{\theequation}{S\arabic{equation}}
\renewcommand{\thefigure}{S\arabic{figure}}
\renewcommand{\bibnumfmt}[1]{[S#1]}


\vspace{8mm}

This supplement material contains five parts: I. A detailed derivation for the connection between the scattering matrix ($S$-Matrix) and the effective Hamiltonian; II. A detailed discussion of the multi-cavity cases under the resonance condition; III. A detailed discussion about the effect of the fabrication-induced disorder; IV. A detailed discussion about the effect of non-mirror configuration; V. A detailed discussion about the effect of the external loss channels. VI. A detailed discussion about two configurations of the waveguide direct- and side-coupled to the cavity. VII. A detailed discussion on the waveguide side-coupled to two systems with distinct energy-level structures; VIII. A detailed proof of Eq.(6); IX. Validating the scattering matrix calculation for the single cavity case. X. A detail about the matrix product states (MPS) simulations for the multi-cavity case.

\tableofcontents

\section{Proof for the Connection between the $S$-Matrix and the Effective Hamiltonian}
In this section, we provide a detailed proof to support Eq.\,(2) in the main text, explaining the connection between the $S$-Matrix elements and the effective Hamiltonian in the waveguide-cavity quantum electrodynamics (QED) system under the Markovian approximation.

\subsection{The Input-Output Formalism and the Quantum Causality Relation}
Firstly, the total Hamiltonian describing multiple cavities coupling to a two-mode waveguide reads as
\begin{align}\label{eq1}
	H_{\text{tot}}=H_{\text{sys}}+\sum_{\mu=l,r}\int\dd{\omega}\ \omega b_\mu^\dagger(\omega)b^{}_\mu(\omega)+\sum_{j=1}^N\sum_{\mu=l,r}\sqrt{\frac{\kappa_\mu}{2\pi}}\int\dd{\omega}[b_\mu^\dagger(\omega)a^{}_je^{-i\omega x_j/v_\mu}+a^\dagger_jb^{}_\mu(\omega)e^{i\omega x_j/v_\mu}],
\end{align}
where $H_{\text{sys}}=\sum_{j=1}^{N}[\omega_ca_j^\dagger a_j^{}+\omega_e\sigma_j^\dagger\sigma^{}_j+\text{g}(a_j^\dagger\sigma^{}_j+\sigma_j^\dagger a^{}_j)]$ and $x_j=(j-1)d$.
Then, we select a rotating frame, and the corresponding unitary transformation is represented by $U_0=\exp(iH_0t)$, where
\begin{align}\label{eq2}
	H_0 = \omega_d\sum_{j=1}^{N}[a_j^\dagger a^{}_j + \sigma^\dagger_j\sigma^{}_j]+\omega_d\sum_{\mu=l,r}\int\dd{\omega}b_\mu^\dagger(\omega)b^{}_\mu(\omega).
\end{align}
Thus, the total Hamiltonian under the rotating frame can be written as
\begin{align}\label{eq3}
	\tilde{H}_{\text{tot}}=U^{}_0H_{\text{tot}}U_0^\dagger+i(\partial_tU^{}_0)U_0^\dagger,
\end{align}
and the Heisenberg equations of motion are
\begin{subequations}
	\begin{align}
		\frac{\dd}{\dd t}\tilde{b}_\mu(\omega)&=-i(\omega-\omega_d) \tilde{b}_\mu(\omega)-i\sum_{j=1}^N\sqrt{\frac{\kappa_\mu}{2\pi}}\tilde{a}_je^{-i\omega x_j/v_\mu},\label{eq4a}\\
		\frac{\dd}{\dd t}\tilde{a}_m&=-i[\tilde{a}_m, \tilde{H}_{\text{sys}}]-i\sum_{\mu=l,r}\sqrt{\frac{\kappa_\mu}{2\pi}}\int\dd{\omega}\tilde{b}_\mu(\omega)e^{i\omega x_m/v_\mu},\label{eq4b}
	\end{align}
\end{subequations}
where $\tilde{H}_{\text{sys}}=\sum_{j=1}^{N}[(\omega_c-\omega_d)\tilde{a}_j^\dagger \tilde{a}_j^{}+(\omega_e-\omega_d)\tilde{\sigma}_j^\dagger\tilde{\sigma}_j^{}+\text{g}(\tilde{a}_j^\dagger\tilde{\sigma}_j^{}+\tilde{\sigma}_j^\dagger\tilde{a}_j^{})]$, and the tildes indicate operators after transformation $U_0$ (i.e., in the rotating frame). We then define the input and output operators as 
\begin{subequations}
	\begin{align}
		\tilde{b}_{\mu, \text{in}}(t)&\equiv\int\frac{\dd{\omega}}{\sqrt{2\pi}}\tilde{b}_{\mu}(\omega, t_0)e^{-i(\omega-\omega_d)(t-t_0)},\quad t_0\to-\infty, \label{eq5a}\\
		\tilde{b}_{\mu, \text{out}}(t)&\equiv\int\frac{\dd{\omega}}{\sqrt{2\pi}}\tilde{b}_{\mu}(\omega, t_1)e^{-i(\omega-\omega_d)(t-t_1)},\quad t_1\to+\infty. \label{eq5b}
	\end{align}
\end{subequations}
After multiplying Eq.\,(\ref{eq4a}) by the integration factor $\exp[i(\omega-\omega_d)t]$, we integrate it from an initial time $t_0<t$ to obtain
\begin{align}\label{eq6}
	\tilde{b}_\mu(\omega,t)=\tilde{b}_\mu(\omega,t_0)e^{-i(\omega-\omega_d)(t-t_0)}-i\sum_{j=1}^N\sqrt{\frac{\kappa_\mu}{2\pi}}\int_{t_0}^{t}\dd{t^\prime}\tilde{a}_j(t^\prime)e^{-i(\omega-\omega_d)(t+x_j/v_\mu-t^\prime)}e^{-i\omega_d x_j/v_\mu}. 
\end{align}
Combining with Eq.\,(\ref{eq5a}), we integrate Eq.\,(\ref{eq6}) with respect to $\omega$ and obtain
\begin{align}\label{eq7}
	\int\frac{\dd{\omega}}{\sqrt{2\pi}}\tilde{b}_{\mu}(\omega,t)=\tilde{b}_{\mu,\text{in}}(t)-i\sqrt{\kappa_\mu}\sum_{j=1}^Ne^{-i\omega_d x_j/v_\mu}\int_{t_0}^{t}\dd{t^\prime}\tilde{a}_j(t^\prime)\delta(t+x_j/v_\mu-t^\prime).
\end{align}
Similarly, we integrate Eq.\,(\ref{eq4a}) up to a final time $t_1>t$, and take advantage of Eq.\,(\ref{eq5b}), which results in 
\begin{align}\label{eq8}
	\int\frac{\dd{\omega}}{\sqrt{2\pi}}\tilde{b}_{\mu}(\omega,t)=\tilde{b}_{\mu,\text{out}}(t)+i\sqrt{\kappa_\mu}\sum_{j=1}^Ne^{-i\omega_d x_j/v_\mu}\int_{t}^{t_1}\dd{t^\prime}\tilde{a}_j(t^\prime)\delta(t+x_j/v_\mu-t^\prime).
\end{align}
Under the usual Markovian approximation, i.e., $\abs{\omega_c-\omega_d},\abs{\omega_e-\omega_d},\text{g},\kappa_\mu\ll \abs{v_\mu}/\max\limits_{j,l}\abs{x_j-x_l}$, we have $\tilde{a}_j(t\pm \abs{x_j/v_\mu})\approx \tilde{a}_j(t)$ and $\exp(-i\omega_dx_j/v_\mu)\approx\exp(-i\omega_cx_j/v_\mu)$. As a result, we could derive the following input-output formalism from the above Eqs.\,(\ref{eq7},\ref{eq8}) and obtain
\begin{align}\label{eq9}
	\tilde{b}_{\mu,\text{out}}(t) = \tilde{b}_{\mu, \text{in}}(t) - i\sqrt{\kappa_\mu}\sum_{j=1}^Ne^{-i\omega_c x_j/v_\mu}\tilde{a}_j(t).
\end{align}
Furthermore, substituting Eq.\,(\ref{eq7}) and Eq.\,(\ref{eq8}) into Eq.\,(\ref{eq4b}) results in
\begin{subequations}
	\begin{align}
		\frac{\dd}{\dd t}\tilde{a}_m(t)=&-i[\tilde{a}_m(t), \tilde{H}_{\text{sys}}]-i\sum_{\mu=l,r}\sqrt{\frac{\kappa_\mu}{2\pi}}\tilde{b}_{\mu, \text{in}}(t)e^{i\omega_d x_m/v_\mu}-\frac{\kappa_l+\kappa_r}{2}\tilde{a}_m(t)\nonumber\\
		&-\sum_{n=1}^{m-1}\kappa_re^{-i\omega_d(x_n-x_m)/v_r}\tilde{a}_n(t)-\sum_{n=m+1}^{N}\kappa_le^{-i\omega_d(x_n-x_m)/v_l}\tilde{a}_n(t),\label{eq10a}\\
		\frac{\dd}{\dd t}\tilde{a}_m(t)=&-i[\tilde{a}_m(t), \tilde{H}_{\text{sys}}]-i\sum_{\mu=l,r}\sqrt{\frac{\kappa_\mu}{2\pi}}\tilde{b}_{\mu, \text{out}}(t)e^{i\omega_d x_m/v_\mu}+\frac{\kappa_l+\kappa_r}{2}\tilde{a}_m(t)\nonumber\\
		&+\sum_{n=1}^{m-1}\kappa_re^{-i\omega_d(x_n-x_m)/v_r}\tilde{a}_n(t)+\sum_{n=m+1}^{N}\kappa_le^{-i\omega_d(x_n-x_m)/v_l}\tilde{a}_n(t).\label{eq10b}
	\end{align}
\end{subequations}
Based on Eqs.\,(\ref{eq10a},\ref{eq10b}), we could obtain a crucial property, namely quantum causality relation, which states that
\begin{subequations}
	\begin{align}
		[\tilde{a}_m(t), \tilde{b}_{\mu, \text{in}}(t^\prime)] = [\tilde{a}^\dagger_m(t), \tilde{b}_{\mu, \text{in}}(t^\prime)] = [\tilde{a}_m(t), \tilde{b}^\dagger_{\mu, \text{in}}(t^\prime)]= [\tilde{a}^\dagger_m(t), \tilde{b}^\dagger_{\mu, \text{in}}(t^\prime)]=0, \text{for}\ t\le t^\prime,\label{eq11a}\\
		[\tilde{a}_m(t), \tilde{b}_{\mu, \text{out}}(t^\prime)] = [\tilde{a}^\dagger_m(t), \tilde{b}_{\mu, \text{out}}(t^\prime)] = [\tilde{a}_m(t), \tilde{b}^\dagger_{\mu, \text{out}}(t^\prime)]= [\tilde{a}^\dagger_m(t), \tilde{b}^\dagger_{\mu, \text{out}}(t^\prime)]=0, \text{for}\ t\ge t^\prime.\label{eq11b}
	\end{align}
\end{subequations}
\subsection{The Derivation of the Multi-Photon $S$ Matrix and the Effective Hamiltonian}\label{IB}
Here, we consider such a scattering process, i.e., incoming $n$ photons from the left-side input port being scattered into the right-side output port, and the process can be described by $n$-photon $S$-matrix elements. Combining with the definitions of input and output operators (\ref{eq5a}, \ref{eq5b}), the $n$-photon $S$-matrix elements can be written as
\begin{align}\label{eq12}
	\tilde{S}_{p_1 \ldots p_n;k_1\ldots k_n}^{rr}=\left[\prod_{l=1}^{n}\frac{\dd{t_l^\prime}}{\sqrt{2\pi}}e^{i(p_l-\omega_d)t_l^\prime}\prod_{j=1}^{n}\frac{\dd{t^{}_j}}{\sqrt{2\pi}}e^{-i(k_j-\omega_d)t^{}_j} \right]\langle0|\prod_{l=1}^{n}\tilde{b}^{}_{r,\text{out}}(t^\prime_l)\prod_{j=1}^{n}\tilde{b}^\dagger_{r,\text{in}}(t^{}_j)|0\rangle,
\end{align}
and we need to compute the time-domain $n$-photon $S$ matrix,
\begin{align}\label{eq13}
	\tilde{S}^{rr}_{t_1^\prime\ldots t_n^\prime;t_1^{}\ldots t_n^{}}\equiv \langle0|\prod_{l=1}^{n}\tilde{b}^{}_{r,\text{out}}(t_l^\prime)\prod_{j=1}^{n}\tilde{b}^\dagger_{r,\text{in}}(t^{}_j)|0\rangle.
\end{align}
Here for simplicity, let us set $\tilde{A}_{\mu,\omega}(t)\equiv\sqrt{\kappa_\mu}\sum_{j=1}^Ne^{-i\omega x_j/v_\mu}\tilde{a}_j(t)$, where $\mu\in\{l,r\}$. By utilizing the input-output relation Eq.\,(\ref{eq9}) and the quantum causality relation Eqs.\,(\ref{eq11a},\ref{eq11b}), we can simplify the computation of the $S$ matrix. This simplification transforms the computation into that of the Green's function, which solely consists of the operator $\tilde{A}_{r,\omega_c}$\,\cite{PhysRevA.91.043845}. Consequently, we focus on evaluating the time-ordered $2M$-point Green's function
\begin{align}\label{eq14}
	\tilde{G}^{rr}(t^\prime_1,\ldots, t^\prime_M;t^{}_1,\ldots, t^{}_M)\equiv(-1)^M\langle0|\mathcal{T}[\tilde{A}^{}_{r,\omega_c}(t_1^\prime)\cdots \tilde{A}^{}_{r,\omega_c}(t_M^\prime)\tilde{A}^\dagger_{r,\omega_c}(t^{}_1)\cdots \tilde{A}^\dagger_{r,\omega_c}(t^{}_M)]|0\rangle,
\end{align}
where $\mathcal{T}$ is a time-ordered symbol. In order to prove the Green's functions can be computed by using the effective Hamiltonian, we start by the path integral formulation of the Green's function, i.e.,
\begin{align}\label{eq15}
	\tilde{G}^{rr}(t_1^\prime,\ldots,t_M^\prime;t_1^{},\ldots,t_M^{})=(-1)^M\frac{\int\mathcal{D}[\{\tilde{b}^{}_{\mu}(\omega),\tilde{b}^*_{\mu}(\omega)\},\{\tilde{a}^{}_j,\tilde{a}^*_j\}]\prod_{j=1}^M\tilde{A}^{}_{r,\omega_c}(t^\prime_j)\prod_{l=1}^M\tilde{A}_{r,\omega_c}^*(t^{}_l)e^{i\int\dd{t}\tilde{\mathcal{L}}}}{\int\mathcal{D}[\{\tilde{b}^{}_{\mu}(\omega),\tilde{b}^*_{\mu}(\omega)\},\{\tilde{a}^{}_j,\tilde{a}^*_j\}]e^{i\int\dd{t}\tilde{\mathcal{L}}}},
\end{align}
where
\begin{align}\label{eq16}
	\tilde{\mathcal{L}}=\sum_{\mu=l,r}\int\dd{\omega}\{\tilde{b}_{\mu}^*(\omega)[i\partial_t-(\omega-\omega_d)]\tilde{b}^{}_\mu(\omega)-[\tilde{b}_\mu^*(\omega)\tilde{A}^{}_{\mu,\omega}(t)+\tilde{A}_{\mu,\omega}\tilde{b}^{}_\mu(\omega)]/\sqrt{2\pi}\}+\tilde{\mathcal{L}}_{\text{sys}}
\end{align}
is the Lagrangian associated with the total Hamiltonian (\ref{eq3}) and $\tilde{\mathcal{L}}_{\text{sys}}$ is the local system's Lagrangian obtained by Legendre transformation on the system's Hamiltonian $\tilde{H}_{\text{sys}}$. Meanwhile, we define a symbol $\mathcal{A}=\partial_t+i(\omega-\omega_d)$ and introduce the propagator of the free waveguide photon
\begin{align}\label{eq17}
	G_{\omega}^{(0)}(t-t^\prime)\equiv\int\frac{\dd{k}}{2\pi}e^{-ik(t-t^\prime)}\frac{i}{k-\omega+i0^+}=e^{-i\omega(t-t^\prime)}\Theta(t-t^\prime),
\end{align}
where $\Theta(t)$ is the step function. Notably, the propagator is the inverse of $\mathcal{A}$, i.e., $\mathcal{A}G_{\omega-\omega_d}^{(0)}(t-t^\prime)=\delta(t-t^\prime)$. 

Finally, we integrate out the waveguide degrees of freedom
\begin{align}
	\int\mathcal{D}[\{\tilde{b}^{}_{\mu}(\omega),\tilde{b}^*_{\mu}(\omega)\}]e^{i\int\dd{t}\tilde{\mathcal{L}}}&=e^{i\int\dd{t}\tilde{\mathcal{L}}_{\text{sys}}}\int\mathcal{D}[\{\tilde{b}^{}_{\mu}(\omega),\tilde{b}^*_{\mu}(\omega)\}]e^{-\sum_{\mu=l,r}\int\dd{t}\int\dd{\omega}[\tilde{b}_\mu^*(\omega)\mathcal{A}\tilde{b}^{}_\mu(\omega)+i\frac{\tilde{b}_\mu^*(\omega)\tilde{A}^{}_{\mu,\omega}(t)+\tilde{A}^*_{\mu,\omega}(t)\tilde{b}^{}_\mu(\omega)}{\sqrt{2\pi}}]}\nonumber\\
	&=\mathscr{N}e^{i\int\dd{t}\tilde{\mathcal{L}}_{\text{sys}}-\sum_{\mu=l,r}\int\dd{t}\int\dd{t^\prime}\int\dd{\omega}\tilde{A}^{}_{\mu,\omega}(t)G_{\omega-\omega_d}^{(0)}(t-t^\prime)\tilde{A}^{}_{\mu,\omega}(t^\prime)/(2\pi)}\nonumber\\
	&=\mathscr{N}e^{i\int\dd{t}\tilde{\mathcal{L}}_{\text{sys}}-\sum_{\mu=l,r}\int\dd{t}\int\dd{t^\prime} \kappa_\mu\sum_{j,k=1}^{N}\tilde{a}_j^*(t)\tilde{a}^{}_k(t^\prime)e^{i\omega_d(x_j-x_k)/v_\mu}\delta(t-t^\prime-\frac{x_j-x_k}{v_\mu})\Theta(t-t^\prime)}\nonumber\\
	&=\mathscr{N}e^{i\int\dd{t}\tilde{\mathcal{L}}_{\text{sys}}-\int\dd{t}\sum_{\mu=l,r} \kappa_\mu\sum_{j,k=1}^{N}\tilde{a}_j^*(t)\tilde{a}^{}_k(t)e^{-i\omega_c(x_k-x_j)/v_\mu}\Theta(\frac{x_j-x_k}{v_\mu})}\nonumber\\
	&=\mathscr{N}e^{i\int\dd{t}\tilde{\mathcal{L}}_{\text{eff}}},\label{eq18}
\end{align}
where $\mathscr{N}$ is the constant coefficient from the Gaussian functional integration. In the fourth step, we use the Markovian approximation. In the last step, the effective Lagrangian can be written as 
\begin{align}\label{eq19}
	\tilde{\mathcal{L}}_{\text{eff}}=\tilde{\mathcal{L}}_{\text{sys}}+i\sum_{j=1}^N\frac{\kappa_l+\kappa_r}{2}\tilde{a}_j^*\tilde{a}_j^{}+i\sum_{j>k}^Ne^{i\omega_c(x_j-x_k)/v_r}[\kappa_r\tilde{a}_j^*\tilde{a}^{}_k+\kappa_l\tilde{a}_k^*\tilde{a}^{}_j],
\end{align}
and the corresponding effective Hamiltonian is obtained from the effective Lagrangian by Legendre transformation, i.e.,
\begin{align}\label{eq20}
	\tilde{H}_{\text{eff}}=\tilde{H}_{\text{sys}}-i\sum_{j=1}^N\frac{\kappa_l+\kappa_r}{2}\tilde{a}_j^\dagger\tilde{a}^{}_j-i\sum_{j>k}^Ne^{i\omega_c(x_j-x_k)/v_r}[\kappa_r\tilde{a}_j^\dagger\tilde{a}^{}_k+\kappa_l\tilde{a}_k^\dagger\tilde{a}^{}_j].
\end{align}
As a result, Eq.\,(\ref{eq15}) is simplified further as
\begin{align}\label{eq21}
	\tilde{G}^{rr}(t_1^\prime,\ldots,t_M^\prime;t_1^{},\ldots,t_M^{})&=(-1)^M\frac{\int\mathcal{D}[\{\tilde{a}^{}_j,\tilde{a}^*_j\}]\prod_{j=1}^M\tilde{A}^{}_{r,\omega_c}(t^\prime_j)\prod_{l=1}^M\tilde{A}_{r,\omega_c}^*(t_l)e^{i\int\dd{t}\tilde{\mathcal{L}}_{\text{eff}}}}{\int\mathcal{D}[\{\tilde{a}^{}_j,\tilde{a}^*_j\}]e^{i\int\dd{t}\tilde{\mathcal{L}}_{\text{eff}}}}\nonumber\\
	&={G}^{rr}(t_1^\prime,\ldots,t_M^\prime;t_1^{},\ldots,t_M^{})\equiv(-1)^M\langle0|\mathcal{T}[\prod_{j=1}^M{A}^{}_{r,\omega_c}(t_j^\prime)\prod_{l=1}^M{A}^\dagger_{r,\omega_c}(t^{}_l)]|0\rangle
\end{align}
with operators
\begin{align}\label{eq22}
	{A}_{r,\omega_c}(t)=e^{iH_{\text{eff}}t}A_{r,\omega_c}e^{-iH_{\text{eff}}t},\quad {A}^\dagger_{r,\omega_c}(t)=e^{iH_{\text{eff}}t}A^\dagger_{r,\omega_c}e^{-iH_{\text{eff}}t},
\end{align} 
where 
\begin{align}\label{eq23}
	H_{\text{eff}}=H_{\text{sys}}-i\sum_{j=1}^N\frac{\kappa_r+\kappa_l}{2}a_{j}^\dagger a_{j}^{}-i\sum_{j>k}^Ne^{i\phi(j-k)}[\kappa_ra_{j}^\dagger a_{k}^{}+\kappa_la_{k}^\dagger a_{j}^{}],
\end{align}
with $\phi=\omega_cd/v_r$. In the second and last steps of Eq.\,(\ref{eq21}), we have already reverted to the original frame [i.e., Eq.\,(\ref{eq1})] and removed the tildes so as to indicate system's operators after the corresponding transformation. Accordingly, the $S$-matrix elements should also take off the coefficient $\omega_d$ in Eq.\,(\ref{eq12}) when the whole system reverts to the original frame. Thus, we have obtained the effective Hamiltonian Eq.\,(2) in the main text, as shown in Eq.\,(\ref{eq23}), and proved that the computation of the scattering matrix elements are completely depended on the effective Hamiltonian for the system without involving the waveguide photon's degree of freedom.

\section{Discussion of the Multi-Cavity Cases under the Resonance Condition}
To support the output states in the main text and the case of multi-cavity under the resonance condition in the main text, in this section, we provide the detailed derivation for the multi-photon $S$-matrix elements and the second-order equal-time correlation function. For this purpose, we again introduce the scattering process mentioned at Sec.\,\ref{IB} and assume that the frequencies of incoming $n$ photons are all $\omega_d$. Therefore, the input state can be defined as $|\Psi_{\text{in}}^{(n)}\rangle^r_{\omega_d}=[b^{\dagger n}_r(\omega_d)/\sqrt{n!}]|0\rangle_r$, and the output state being scattered into the right-side waveguide is given by
\begin{align}
	|\Psi_{\text{out}}^{(n)}\rangle^r_{\omega_d}=\int\frac{\prod_{j=1}^n\dd{p_j}}{n!\sqrt{n!}}[\prod_{j=1}^{n}b^\dagger_r(p_j)]|0\rangle\langle0|[\prod_{j=1}^{n}b_r(p_j)]{S}[b_r^{\dagger n}(\omega_d)]|0\rangle_r
	=\int\frac{\prod_{j=1}^n\dd{p_j}}{n!\sqrt{n!}}{S}^{rr}_{p_1\ldots p_n;\omega_d\ldots\omega_d}[\prod_{j=1}^{n}b^\dagger_r(p_j)]|0\rangle_r.\label{eq24}
\end{align}
When we consider the cases of $n=1$ and $n=2$, Eq.\,(\ref{eq24}) will give the single- and two-photon output states of the main text. Besides, for the single-photon $S$ matrix, we have
\begin{align}\label{eq25}
	{S}_{p;\omega_d}^{rr}=\delta(p-\omega_d)[1+O^r_{0,1}\mathcal{K}^{-1}_{\omega_d}(1)O_{0,1}^{r\dagger}],
\end{align}
where $\mathcal{K}_\omega(n)=-i[H_{\text{eff}}^{(n)}-\omega]$. Here, the projection of the effective Hamiltonian on the $n$-th excitation subspace is denoted as $H_{\text{eff}}^{(n)}$, and the projection of the linear combination $A_{r,\omega_c}$ of annihilation operators onto the direct sum of the $n$-th and $(n+1)$-th excitation subspace is denoted as $O^{r}_{n, n+1}$. Similarly, the two-photon $S$-matrix elements can be written as
\begin{align}\label{eq26}
	{S}_{p_1p_2;\omega_d\omega_d}^{rr}=2{S}^{rr}_{p_1;\omega_d}{S}^{rr}_{p_2;\omega_d}+\delta(2\omega_d-p_1-p_2){T}^{}_{p_1p_2;\omega_d\omega_d}
\end{align}
with the intrinsic two-photon correlation term
\begin{align}\label{eq27}
	{T}_{p_1p_2;\omega_d\omega_d}=\frac{1}{\pi}O^r_{0,1}[\mathcal{K}_{2\omega_d-p_1}^{-1}(1)+\mathcal{K}_{2\omega_d-p_2}^{-1}(1)][O^r_{1,2}\mathcal{K}^{-1}_{2\omega_d}(2)O_{1,2}^{r\dagger}
	-O^r_{0,1}\mathcal{K}^{-1}_{\omega_d}(1)O_{0,1}^{r\dagger}]\mathcal{K}^{-1}_{\omega_d}(1)O_{0,1}^{r\dagger}.
\end{align}

\subsection{Even Number of Cavities}\label{IIA}
Here, we consider the case of $N=2j$, i.e., an even number of cavities, and assume $\omega_c=\omega_e=\omega_d=\omega$, which is known as the resonance condition. Note that we only consider a mirror configuration (i.e., $\phi=2\pi$) in this section. Firstly, we analyze the case of $j=1$ to simplify the proof. Thus, in the single- and two-excitation number subspace, the effective Hamiltonian $H_{\text{eff}}$ has
\begin{align}\label{eq28}
	H_{\text{eff}}^{(1)}=\left[ \begin{matrix}
		\omega-\frac{i\kappa}{2}&		\text{g}&		-i\kappa _l&		0\\
		\text{g}&		\omega&		0&		0\\
		-i\kappa _r&		0&		\omega-\frac{i\kappa}{2}&		\text{g}\\
		0&		0&		\text{g}&		\omega\\
	\end{matrix} \right],H_{\text{eff}}^{(2)}=\left[ \begin{matrix}
		2\omega-i\kappa&		\text{g}\sqrt{2}&		-i\kappa_l\sqrt{2}&		0 & 0 & 0 & 0 & 0\\
		\text{g}\sqrt{2}&		2\omega-\frac{i\kappa}{2}&		0&		0 & -i\kappa_l & 0 & 0 & 0\\
		-i\kappa_r\sqrt{2}&		0&		2\omega-i\kappa&		\text{g} & \text{g} & 0 &-i\kappa_l\sqrt{2} & 0\\
		0&		0&		\text{g}&		2\omega-\frac{i\kappa}{2} & 0 &  \text{g} & 0 & -i\kappa_l\\
		0 & -i\kappa_r & \text{g} & 0 & 2\omega-\frac{i\kappa}{2} & \text{g} & 0 & 0\\
		0 & 0 & 0 & \text{g} & \text{g} & 2\omega & 0 & 0\\
		0 & 0 & -i\kappa_r\sqrt{2} & 0 & 0 & 0 & 2\omega-i\kappa & \text{g}\sqrt{2}\\
		0& 0 & 0 &-i\kappa_r & 0 & 0 & \text{g}\sqrt{2} & 2\omega-\frac{i\kappa}{2}
	\end{matrix} \right],
\end{align}
and the linear combination $A_{r,\omega}$ can be written as
\begin{align}\label{eq29}
	\quad O^r_{0,1}=\sqrt{\kappa_r}\left[ \begin{matrix}
		1&		0&		1&		0
	\end{matrix} \right],\quad
	O^r_{1,2}=\sqrt{\kappa_r}\left[ \begin{matrix}
		\sqrt{2}&		0&		1&		0&		0&		0&		0&		0 \\ 
		0	&		1&		0&		0&		1&		0&		0&		0 \\ 
		0	&		0&		1&		0&		0&		0&	\sqrt{2}&	0 \\ 
		0	&		0&		0&		1&		0&		0&		0&		1  
	\end{matrix} \right].
\end{align}
Notably, we find 
\begin{align}\label{eq30}
	\mathcal{K}_{\omega}^{-1}(1)O_{0,1}^{r\dagger}=\frac{i\sqrt{\kappa_r}}{\text{g}}\left[\begin{matrix}
		0\\1\\0\\1
	\end{matrix}\right],\quad  O^r_{1,2}\mathcal{K}_{2\omega}^{-1}(2)O_{1,2}^{r\dagger}=-\frac{\kappa_r}{\kappa}\left[\begin{matrix}
	1 & 0 & 1 & 0\\
	0 & 1 & 0 & -1\\
	1 & 0 & 1 & 0\\
	0 & -1 & 0 & 1
\end{matrix}\right].
\end{align}
Therefore, according to Eq.\,(\ref{eq25}) and Eq.\,(\ref{eq26}), the single- and two-photon $S$ matrices have
\begin{align}\label{eq31}
	{S}_{p;\omega}^{rr}=\delta(p-\omega),\ {S}^{rr}_{p_1p_2;\omega\omega}=2\delta(p_1-\omega)\delta(p_2-\omega).
\end{align}
Similarly, the $n$-photon $S$ matrix also has the same property, i.e.,
\begin{align}\label{eq32}
	{S}^{rr}_{p_1\ldots p_n;\omega\ldots\omega}=n!\delta(p_1-\omega)\delta(p_2-\omega)\cdots\delta(p_n-\omega).
\end{align}
When we consider a coherent state with amplitude $\eta$ entering into the waveguide from the left side channel, the output state has
\begin{align}
	|\psi_{\text{out}}\rangle={S}|\psi_{\text{in}}\rangle&=\mathcal{N}\sum_{n=0}^{\infty}\frac{\eta^n}{\sqrt{n!}}|\Psi_{\text{out}}^{(n)}\rangle^{r}_{\omega_d}|0\rangle_l|0\rangle_s=\mathcal{N}\sum_{n=0}^{\infty}\frac{\eta^n}{n!}\int\frac{\prod_{j=1}^n\dd{p_j}}{n!}{S}^{rr}_{p_1\ldots p_n;\omega\ldots\omega}[\prod_{j=1}^{n}b^\dagger_r(p_j)]|0\rangle_r|0\rangle_l|0\rangle_s\nonumber\\
	&=\mathcal{N}\sum_{n=0}^{\infty}\frac{\eta^n}{n!}b^{\dagger n}_r(\omega)|0\rangle_r|0\rangle_l|0\rangle_s=|\psi_{\text{in}}\rangle,\label{eq33}
\end{align}
implying that the system does not change the incident coherent state when $N=2$. Similarly, for $j>1$, we can treat the scattering process in the whole system as $j$ cascade processes, and then the output state is still a coherent state.
\subsection{Odd Number of Cavities}\label{IIB}
Here, we consider the case of $N=2j+1\ (j\ge1)$ and use the resonance condition. In the single-excitation number subspace, the effective Hamiltonian could be written as a block-matrix, i.e.,
\begin{align}\label{eq34}
	{H}_{2j+1, \text{eff}}^{(1)}=\left[\begin{matrix}
	{H}_{2j, \text{eff}}^{(1)} & B_{2j,l}^T\\
	B_{2j,r}&H_{1,\text{eff}}^{(1)}
	\end{matrix}\right],
\end{align}
where $H_{2j,\text{eff}}^{(1)}$ represents the effective Hamiltonian of $N=2j$ in the single-excitation number subspace, i.e., the first term of Eq.\,(\ref{eq28}), and 
\begin{align}\label{eq35}
	B_{k, \mu}=-i\kappa_\mu\left[\begin{matrix}
		Z_1&Z_1&\cdots &Z_1
	\end{matrix}\right]_{1\times k},\ H_{1,\text{eff}}^{(1)}=\left[\begin{matrix}
	\omega-i\kappa/2&\text{g}\\
	\text{g}&\omega
\end{matrix}\right],\text{ with }Z_1=\left[\begin{matrix}
1&0\\
0&0
\end{matrix}\right].
\end{align}
Therefore, we have
\begin{align}\label{eq36}
	\mathcal{K}_{\omega}^{-1}(1)=i\left[\begin{matrix}
		Q^{-1}&-Q^{-1}B_{2j,l}^T\mathcal{H}_1^{-1}\\
		-\mathcal{H}_1^{-1}B_{2j,r}Q^{-1}&\mathcal{H}_1^{-1}+\mathcal{H}_1^{-1}B_{2j,r}Q^{-1}B_{2j,r}^T\mathcal{H}_1^{-1}
	\end{matrix}\right],
\end{align}
where $Q=\mathcal{H}_{2j}-B_{2j,l}^T\mathcal{H}_1^{-1}B_{2j,r}$ and $\mathcal{H}_k=H_{k,\text{eff}}^{(1)}-\omega$. On the one hand, we find
\begin{align}\label{eq37}
	B_{2j,l}^T\mathcal{H}_1^{-1}B_{2j,r}=0,\ B_{2j,r}Q^{-1}B_{2j,l}^T=0.
\end{align}
On the other hand, the linear combination $A_{r,\omega}$ also could be written as block-matrix, i.e., 
\begin{align}\label{eq38}
	O^r_{0,1}=\sqrt{\kappa_r}\left[\begin{matrix}
		X_1&X_1&\cdots&X_1
	\end{matrix}\right]_{1\times (2j+1)}, \text{ with } X_1=\left[\begin{matrix}
	1&0
\end{matrix}\right],
\end{align}
and thus we have
\begin{align}\label{eq39}
	\mathcal{K}_\omega^{-1}(1)O_{0,1}^{r\dagger}=\frac{i\sqrt{\kappa_r}}{\text{g}}\left[\begin{matrix}
	X_2&X_2&\cdots&X_2
	\end{matrix}\right]_{1\times (2j+1)}^T, \text{ with } X_2=\left[\begin{matrix}
		0&1
\end{matrix}\right].
\end{align}
Due to $X_1X_2^T=0$ and $P_{1}^{N}\propto O_{0,1}^r\mathcal{K}_{\omega}^{-1}(1)O_{0,1}^{r\dagger}$, the single-photon probability amplitude is equal to zero. Similarly, in two-excitation number subspace, we use the same steps above, and we find
\begin{align}\label{eq40}
	O^r_{1,2}\mathcal{K}_{2\omega}^{-1}(2)O_{1,2}^{r\dagger}=\frac{i}{1+\alpha}\left[\begin{matrix}
		F_+ & 0 & C_+\\
		0 & V & 0\\
		C_- & 0 & F_-
	\end{matrix}\right],
\end{align}
where
\begin{subequations}
	\begin{align}
		F_{\pm}&=\frac{1}{\zeta}\mqty[
		i(1-\alpha)^2(1+\alpha^2)+4i(\text{g}/\kappa)^2(1+\alpha)^4&\mp4(\text{g}/\kappa)\alpha(1-\alpha^2)\\
		\pm4(\text{g}/\kappa)\alpha(1-\alpha^2) & 4i(\text{g}/\kappa)^2(1+\alpha)^2(1+\alpha^2)+i(1-\alpha^2)^2
		],\label{eq41a}\\
		C_{\pm}&=\frac{1}{\zeta}\mqty[
		i(1-\alpha)^2(1+\alpha^2)\pm4i(\text{g}/\kappa)^2(1+\alpha)^2(1-\alpha^2) & \mp4(\text{g}/\kappa)\alpha^{1\pm 1}(1-\alpha^2)\\
		\mp4(\text{g}/\kappa)\alpha^{1\pm 1}(1-\alpha^2)&-4i(\text{g}/\kappa)^2(1+\alpha)^2(1+\alpha^2)\mp i(1-\alpha)^2(1-\alpha^2)
		],\label{eq41b}\\
		V&=i\mathscr{D}\{[\underbrace{Z_2,\cdots,Z_2}_{j-1},Z_1,\underbrace{Z_2,\cdots,Z_2}_{j-1}]\}+i\bar{\mathscr{D}}\{[\underbrace{Z_3,\cdots,Z_3}_{j-1},Z_1,\underbrace{Z_3,\cdots,Z_3}_{j-1}]\},\label{eq41c}
	\end{align}
\end{subequations}
with
\begin{align}\label{eq42}
	Z_2=\left[\begin{matrix}
		1&0\\
		0&1
	\end{matrix}\right],\ Z_3=\left[\begin{matrix}
		1&0\\
		0&-1
\end{matrix}\right].
\end{align}
Here $\mathscr{D}\{\cdot\}$ ($\bar{\mathscr{D}}\{\cdot\}$) of a vector \textbf{v} is a diagonal (back-diagonal) matrix with elements of \textbf{v} on the diagonal (back-diagonal), $\zeta=(1+\alpha^2)[4(\text{g}/\kappa)^2(1+\alpha)^2+(1-\alpha)^2]$, and $\alpha=\kappa_l/\kappa_r$ denoting as the chirality. Due to $X_1Z_sX_2^T=0$ for $s\in\{1,2,3\}$, the two-photon probability amplitude has
\begin{align}
	P_2^{N}\times e^{2i\omega t}&= O^r_{0,1}O^r_{1,2}\mathcal{K}_{2\omega}^{-1}(2)O_{1,2}^{r\dagger}\mathcal{K}_\omega^{-1}(1)O_{0,1}^{r\dagger}
	=\frac{-X_1\sum_{q=\pm}(F_q+C_q)X_2^T}{(g/\kappa)(1+\alpha)^2}=-\frac{4(1-\alpha)^2/(1+\alpha^2)}{\left[4(\text{g}/\kappa)^2(1+\alpha)^2+(1-\alpha)^2\right]}\label{eq43}
\end{align}
corresponding to Eq.\,(10) in the main text. For the sake of convenience, we define a time-independent $n$-photon probability amplitude $P_n^N(0)\equiv P_n^N\exp(in\omega_d t)$. When the magnitude of coherent amplitude approaches to zero, i.e., $\abs{\eta}\to0$, the second-order equal-time correlation function is computed as
\begin{align}\label{eq44}
	g_{rr}^{(2)}(0)=\frac{\abs{1+2P^{N}_1(0)+P^{N}_2(0)}^2}{\abs{1+P^{N}_1(0)}^4}=\abs{\frac{4(\text{g}/\kappa)^2(1+\alpha)^2(1+\alpha^2)-(1-\alpha)^2(3-\alpha^2)}{4(\text{g}/\kappa)^2(1+\alpha)^2(1+\alpha^2)+(1-\alpha)^2(1+\alpha^2)}}^2,
\end{align}
and the analytical parameters condition for the perfect photon blockade (PB) region is
\begin{align}\label{eq45}
	\text{g}/\kappa=\frac{\abs{1-\alpha}}{2(1+\alpha)}\sqrt{\frac{3-\alpha^2}{1+\alpha^2}}\ (0\le\alpha<1).
\end{align}

Finally, we delve into the distinction between the cases of $N=1$ and $N=2j+1\ (j\ge1)$. To analyze this point clearly, we partition the system with $N=2j+1$ into two parts, i.e., $N=N_1+N_2$ with $N_1=2j$ and $N_2=1$. Based on the conclusion of Eq.\,(\ref{eq33}), we observe that two incoming non-interacting photons could freely traverse the first part. However, as these photons proceed through the second part, the two photons that were initially reflected into the left-moving waveguide mode become entangled with each other. Consequently, a second reflection of the two entangled photons by the first part occurs, leading to a scenario where photons can bounces back and forth between the two parts. This accounts for why the system of $N=2j+1$ exhibits more possible scattering paths for photons compared to the system with $N=1$. As a consequence, the optimal parameters condition for perfect PB, described by Eq.\,(\ref{eq45}), is differ from the system of $N=1$. Additionally, in the chiral case (i.e., $\alpha=0$), the distinction is eliminated due to the absence of reflection.

\section{Effects of Disorder}\vspace{-10pt}
\begin{figure}
	\includegraphics[width=14cm]{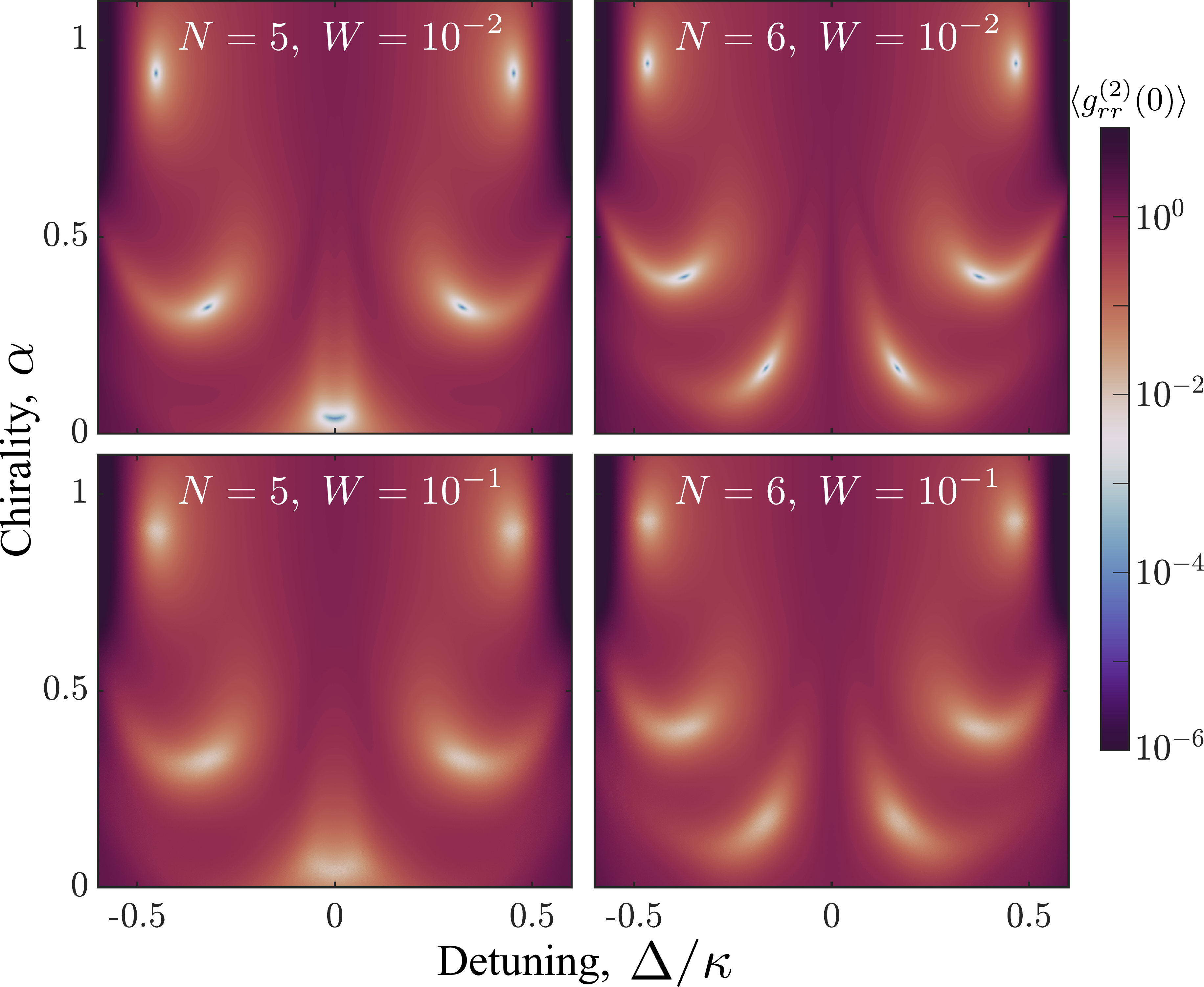}\\
	\caption{The geometric means of equal-time second-order quantum correlation function $\langle g_{rr}^{(2)}(0)\rangle$ versus the chirality $\alpha$ and the detuning $\Delta/\kappa$, for different disorder strength $W$ and the number of cavities $N$. Here, the geometric means computed with a total of $10^3$ instances of disorder. All parameters are the same as in Fig.\,3 of the main text.
	}\label{Fig_Sup6}
\end{figure}
Thus far, we have assumed uniform values for each cavity and atom parameter. In experiments, however, fabrication imperfections and other errors will introduce disorder. Here, we study how disorder in both the cavity and two-level emitter frequencies might affect the perfect PB phenomenon in multi-cavity systems. This discussion is critical, as such fabrication-induced disorder and experimental errors can significantly impact the performance and reliability of the proposed single-photon sources. 

Incorporating the original effective Hamiltonian [see Eq.~(\ref{eq23})], we introduce a random, uniformly distributed frequencies $\epsilon_{\nu,j}/\kappa\in[-W, W]$ for the $j$th cavity ($\nu=c$) and the $j$th emitter ($\nu=e$), where $W$ represents a disorder strength. Consequently, the effective Hamiltonian with disorder can be expressed as
\begin{align}
	H_{\text{eff}}=\sum_{j=1}^{N}[(\omega_c+\epsilon_{c,j}-i\kappa/2)a_j^\dagger a_j^{}+(\omega_e+\epsilon_{e,j})\sigma_j^\dagger\sigma^{}_j+\text{g}(a_j^\dagger\sigma^{}_j+\sigma_j^\dagger a^{}_j)]-i\sum_{j>k}^Ne^{i\phi(j-k)}[\kappa_ra_{j}^\dagger a_{k}^{}+\kappa_la_{k}^\dagger a_{j}^{}].
\end{align}
To estimate the values of $W$ corresponding to strong or weak disorder, a reference value is necessary. As shown in Fig.\,3 of the main text, the detuning corresponding to near-perfect PB lies within the range of $-0.6\kappa$ to $0.6\kappa$. Therefore, with the same parameters (i.e., $\mathrm{g}=0.8\kappa$), we define $W\ll 0.6$ as weak disorder and $W\lesssim0.6$ as strong disorder.

Fig.\,\ref{Fig_Sup6} presents the geometric means of the second-order quantum correlation over $1000$ independent disorder instances for various $N$ and disorder strength $W$, plotted against the detuning $\Delta$ and chirality $\alpha$. In the first row of Fig.\,\ref{Fig_Sup6}, we show the results of weak disorder for $N=5$ and $N=6$. We observe that the disorder shifts the optimal points of perfect PB without eliminating them. As the disorder strength increases, as shown in the second row of Fig.\,\ref{Fig_Sup6}, we find that strong disorder slightly increases the value of the second-order quantum correlation, but the positions of the minima remain close to the perfect PB points in the disorder-free model. The perfect PB phenomenon is thus robust against disorder, as might be given their origin as exact zeros of the total two-photon probability amplitude. The fact that both the $N=5$ and $N=6$ cases show good agreement between Fig.\,\ref{Fig_Sup6} and Figs.\,3(c, f) in the main text indicates that PB does not become more sensitive to disorder with increasing $N$.

\section{Effects of Non-Mirror Configuration}
\begin{figure}
	\includegraphics[width=17cm]{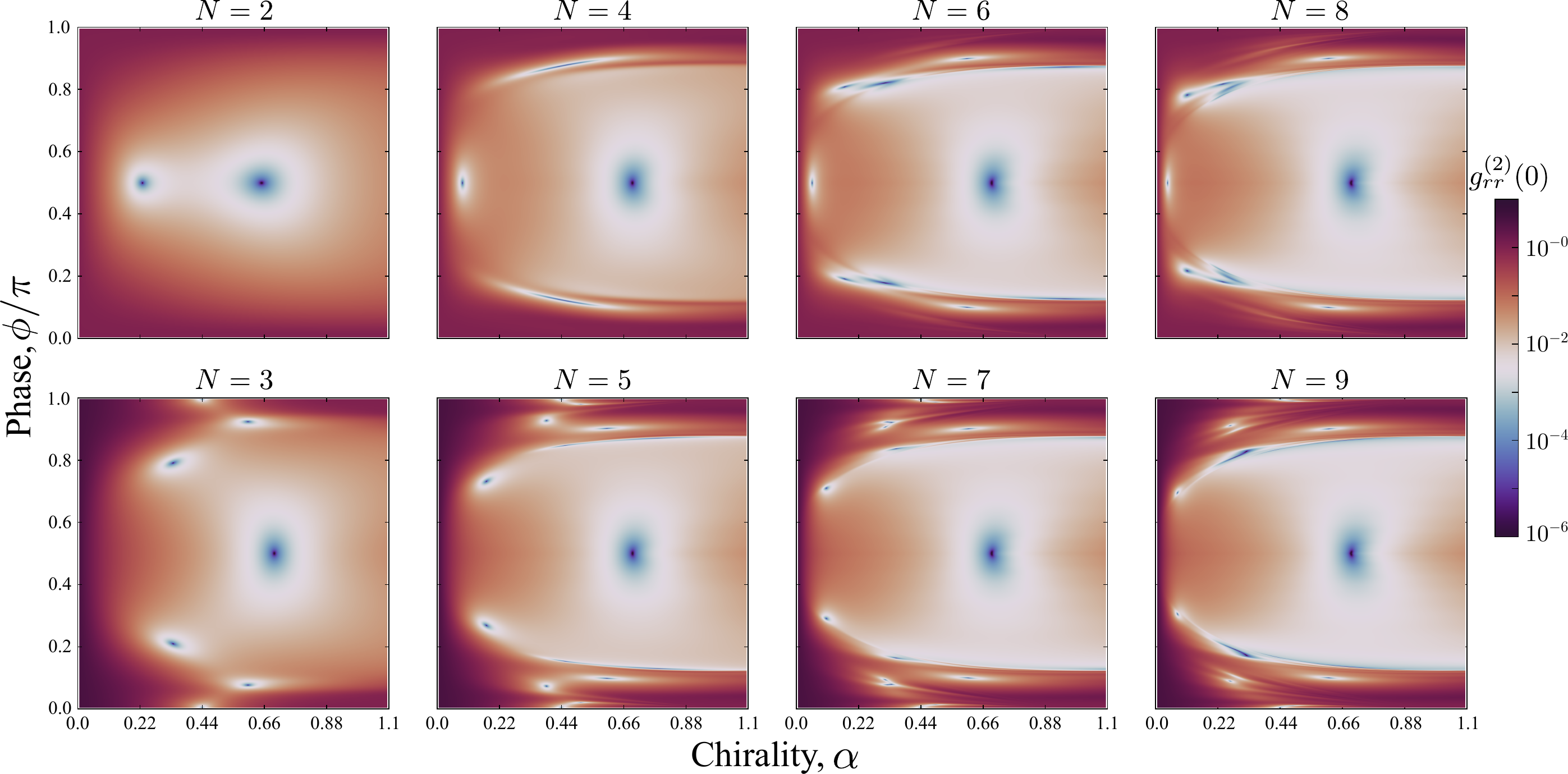}\\
	\caption{The equal-time second-order quantum correlation function $g_{rr}^{(2)}(0)$ versus the chirality $\alpha$ and the phase $\phi/\pi$, for various sizes $N$. The calculation has been performed for $\Delta/\kappa=0$ and $\text{g}=0.8\kappa$.
	}\label{fig1}
\end{figure}
Because we only consider the mirror configuration situation in the main text, in this section, we select a non-mirror configuration situation, i.e., $\phi\neq 2m\pi, m\in\mathbb{N}$, and analyse how the finite phase $\phi=\omega_cd/v_r$, which is gained by light while traveling between the cavities, affects the statistical properties of the output light.

In Fig.\,\ref{fig1}, we show how finite $\phi$ affects the second-order correlation function in the presence of chirality for different $N$. Note that, the correlation function with respect to $\phi=(2m+1)\pi$ is the same as that in the case of $\phi=2m\pi$, where $\phi=m\pi$ corresponds to the Bragg resonances. In the anti-Bragg case, $\phi=\pi/2$, two near-perfect PB points are observed in the system with even number of cavities (upper panels). The first PB point rapidly tends towards the perfect chirality point, characterized by $\alpha_{\text{opt}}\to0$, as the number of cavities increases, while the second PB point exhibits only a tiny change. Similarly, the second PB point is also observed in systems with an odd number cavities (bottom panels), locating at the same chirality position, i.e., $\alpha_{\text{opt}}^{N=2j}=\alpha_{\text{opt}}^{N=2j+1}$. Furthermore, we observe that the correlation function is symmetric with respect to $\phi=\pi/2$, and the near-perfect PB point also exists in other phases, particularly in systems with $N=2j+1$. Additionally, the pronounced antibunching region expands as the number of cavities increases. Undoubtedly, the non-mirror configuration situation also exhibits a significant impact on the statistical properties of the output light, especially in the anti-Bragg resonances.

\section{Effects of External Loss Channels}\vspace{-10pt}
In this section, we provide a detailed discussion about the effect of the external loss channels (non-waveguide modes). For the sake of simplicity, we assume that the atoms and the cavities both are radiate into the different loss channels with annihilation operators $c_j(\omega)$ and $d_j(\omega)$, respectively. The free Hamiltonian and the interaction Hamiltonian for the loss channels coupled to the atom are given by 
\begin{align}
	H_{\text{loss},\mathrm{e}}=\sum_{j=1}^N{\int\dd{\omega}\omega c_j^\dagger(\omega)c_j^{}(\omega)},\quad H_{\text{int},\mathrm{e}}=\sum_{j=1}^N\sqrt{\frac{\gamma_j}{2\pi}}\int\dd{\omega}[c_j^\dagger(\omega)\sigma_j^{}+\sigma_j^\dagger c_j^{}(\omega)],
\end{align}
and the free Hamiltonian and the interaction Hamiltonian for the loss channels coupled to the cavity are given by
\begin{align}
	H_{\text{loss},\mathrm{c}}=\sum_{j=1}^N{\int\dd{\omega}\omega d_j^\dagger(\omega)d_j^{}(\omega)},\quad H_{\text{int},\mathrm{c}}=\sum_{j=1}^N\sqrt{\frac{\kappa_j}{2\pi}}\int\dd{\omega}[d_j^\dagger(\omega)a_j^{}+a_j^\dagger d_j^{}(\omega)].
\end{align}
Here, $\gamma_j$ ($\kappa_j$) denotes as the coupling constant between the $j$-th atom (cavity) and the $j$-th loss channel with mode $c_j$ ($d_j$), and all the coupling constants are set to identical, i.e., $\gamma_j=\gamma_e$ and $\kappa_j=\kappa_e$. Thus, by adding $H_{\text{loss},\mathrm{e}}$, $H_{\text{loss},\mathrm{c}}$, $H_{\text{int},\mathrm{e}}$, and $H_{\text{int},\mathrm{c}}$ in the total Hamiltonian \,(\ref{eq1}) and following the steps of Eqs.\,(\ref{eq1}-\ref{eq23}), we can derive a new effective Hamiltonian, which is given by
\begin{align}\label{eq47}
	H_{\text{eff}}=H_{\text{sys}}-i\sum_{j=1}^N\frac{\gamma_e}{2}\sigma_j^\dagger\sigma_j^{}-i\sum_{j=1}^N\frac{\kappa_r+\kappa_l+\kappa_e}{2}a_{j}^\dagger a_{j}^{}-i\sum_{j>k}^Ne^{i\phi(j-k)}[\kappa_ra_{j}^\dagger a_{k}^{}+\kappa_la_{k}^\dagger a_{j}^{}].
\end{align}
\subsection{The Presence of Atomic Dissipation}
\begin{figure}
	\includegraphics[width=17cm]{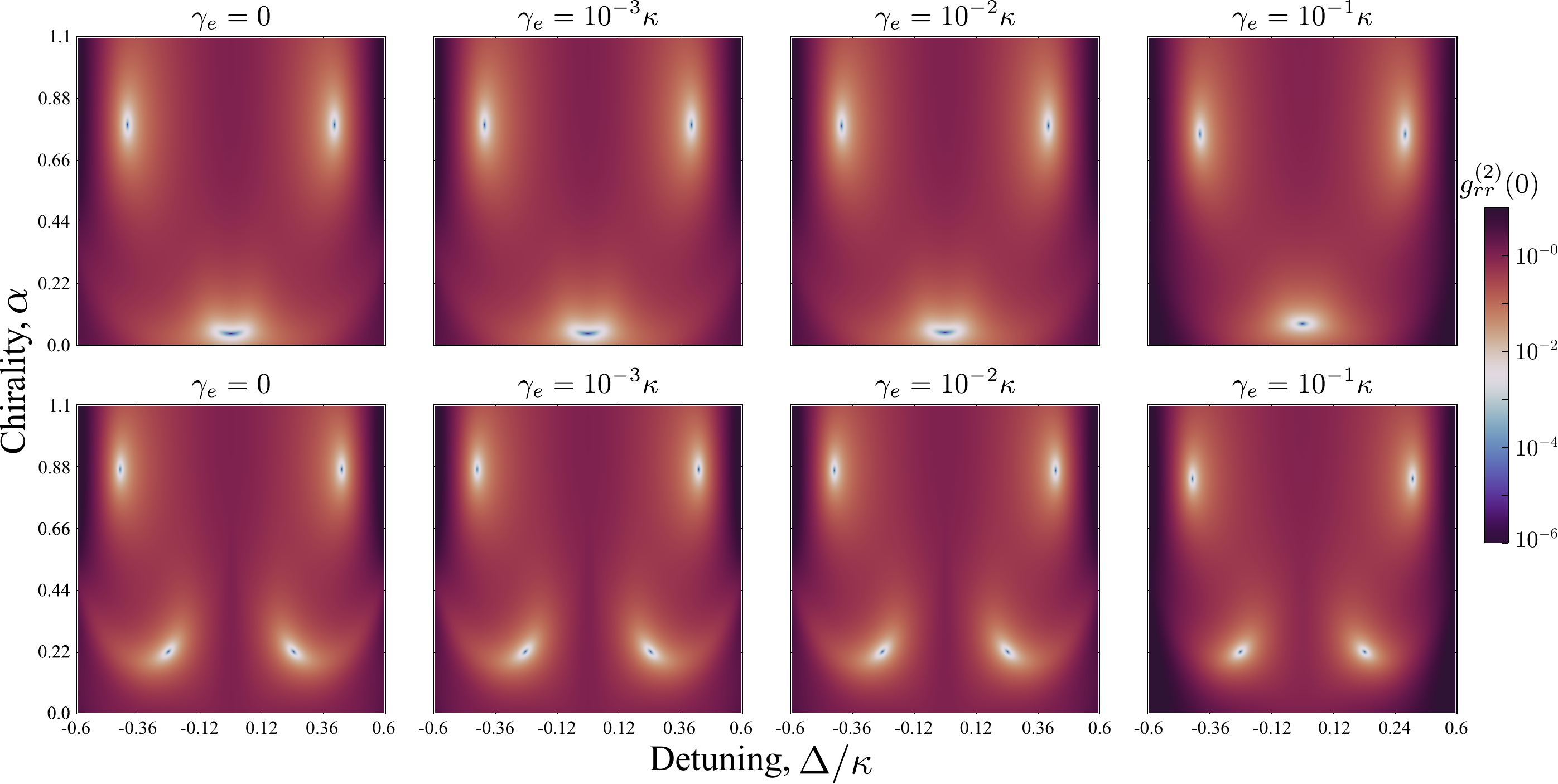}\\
	\caption{The equal-time second-order quantum correlation function $g_{rr}^{(2)}(0)$ versus the chirality $\alpha$ and the detuning $\Delta/\kappa$, for different decay rates $\gamma_e$. In the first line, we consider the system of $N=3$, whereas in the second line, we consider the system of $N=4$. The calculation has been performed for $\phi=2\pi$, $\text{g}=0.8\kappa$, and $\kappa_e=0$.
	}\label{fig2}
\end{figure}
In the main text, the reason we neglect the presence of atomic dissipation is because the small decay rates of the atom does not essentially affect our results. In normal experimental settings, the decay rate of the atoms is significantly smaller compared to that of the cavities. For this purpose, we will show that this assumption in the main text always is valid. Due to the presence of atomic dissipation, it is possible that the incoming photons are scattered into the corresponding loss channels, and the scattering probability is proportional to its coupling constant. For the case of $\gamma_e\ll\kappa$, this scattering probability is negligible compared to the probability of the incoming photons being scattered into the waveguide. As shown in Fig.\,\ref{fig2}, as atomic dissipation increases, the locations of PB points remain almost unchanged for different numbers of cavities. Consequently, for the parameter region of interest to us, the second-order correlation function remains largely unaffected by the scattering into the loss channels, and this conclusion is substantiated by comparing the last three columns (with atomic dissipation) with the first column (without atomic dissipation) in Fig.\,\ref{fig2}. 

\begin{figure}
	\includegraphics[width=17cm]{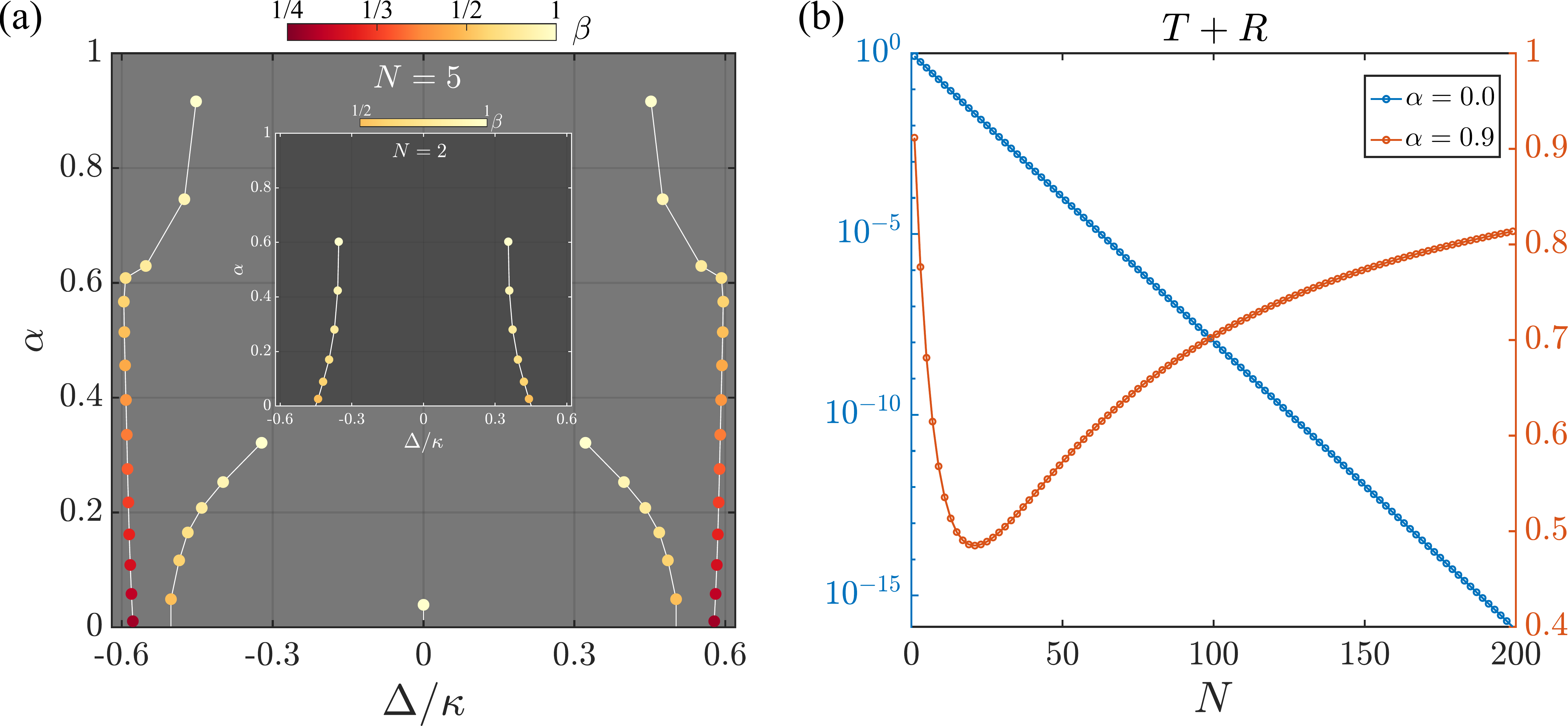}\\
	\caption{Panel (a) illustrates the change in the distribution of perfect PB points (i.e., $g_{rr}^{(2)}(0)=0$) in the parameters space $(\Delta, \alpha)$ as the coupling efficiency $\beta$ decreases, for $N=5$. Panel (b) depicts the single-photon survival rate within the waveguide as a function of the number of cavity $N$. I The inset in panel (a) shows the case of $N=2$. In panel (b), we take $\kappa_e/\kappa=20$ and $\Delta/\kappa=0.2$. The other parameters are given by $\phi=2\pi$, $\text{g}=0.8\kappa$, and $\gamma_e=0$.
	}\label{Fig_Sup5}
\end{figure}
\subsection{The Presence of Cavity's External Dissipation}
Here, we use the symbol $\beta$ to represent the coupling efficiency of the light-matter interfaces between a cavity and a two-mode waveguide. According to Eq.\,(\ref{eq47}), the coupling efficiency is given by $\beta=\kappa/(\kappa+\kappa_e)$. In the main text, we set the coupling efficiency to one (i.e., $\kappa_e=0$), which corresponds to a region of strong coupling to the waveguide. In this subsection, we will investigate whether the main conclusions presented in the main text are influenced under the condition of weak coupling to the waveguide. As the coupling efficiency decreases ($\beta\downarrow$), the perfect PB points overall shift towards increasing chirality ($\alpha\downarrow$) until it disappears, i.e., $\alpha_{\text{ppb}}(\beta_1)<\alpha_{\text{ppb}}(\beta_2)$ for $\beta_1<\beta_2$, where $\alpha_{\text{ppb}}$ denotes the chirality corresponding to the generation of perfect PB, as shown in Fig.\,\ref{Fig_Sup5}(a). Notably, we find that when $\beta=1$, the weaker the chirality $\alpha_{\text{ppb}}$ corresponding to the perfect PB, the smaller the coupling efficiency $\beta$ at which the perfect PB effect completely disappears, as shown in the inset of Fig.\,\ref{Fig_Sup5}(a). This indicates stronger immunity to the dissipation caused by the cavity decaying into non-waveguide loss channels. For the parameters used in Fig.\,\ref{Fig_Sup5}(a), we numerically predict that when $\beta\lesssim0.9569$, the perfect PB at the smallest $\alpha_{\text{ppb}}$ completely disappears; when $\beta\lesssim0.4808$, it disappears at the second smallest $\alpha_{\text{ppb}}$; and when $\beta\lesssim0.2597$, it vanishes at the largest $\alpha_{\text{ppb}}$. In conclusion, at $\beta\lesssim0.2597$, the system will no longer exhibit a perfect PB effect within the parameters space ($\Delta$, $\alpha$).

\section{Waveguide Direct-Coupled and Side-Coupled to a Cavity-Atom System}\label{VI}\vspace{-10pt}
\begin{figure}
	\includegraphics[width=17cm]{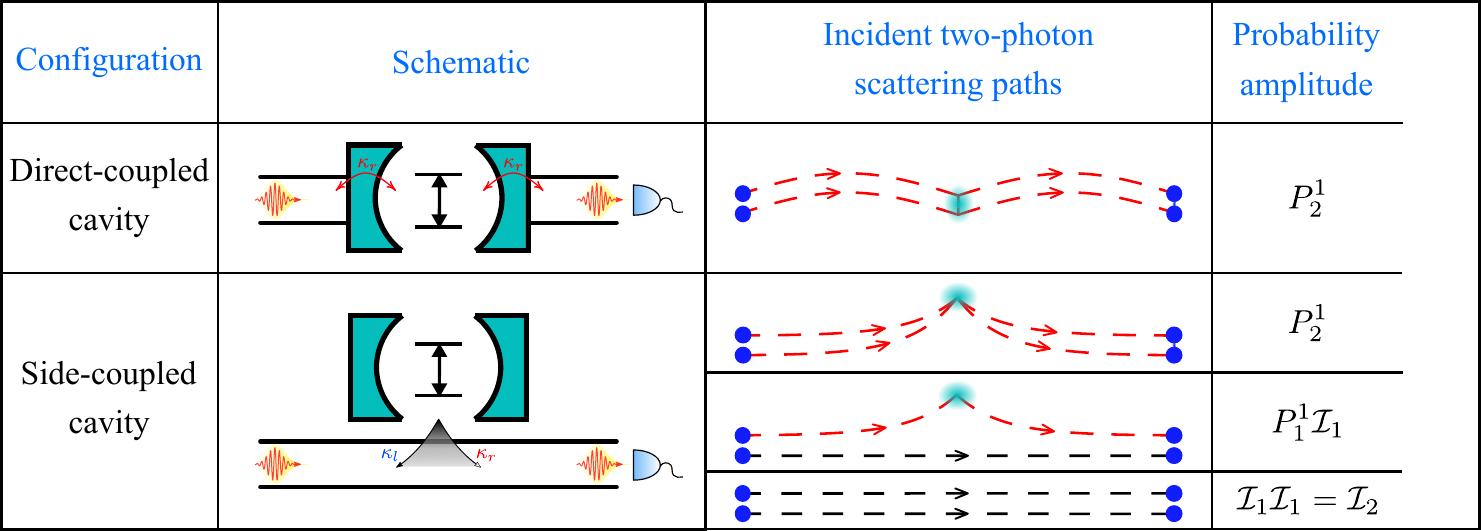}\\
	\caption{The specific schematics, incident two-photon scattering paths, and corresponding probability amplitudes for the two configurations: the direct-coupled cavity and side-coupled cavity.
	}\label{Fig_Sup7}
\end{figure}
In this section, we will discuss these two configurations in detail through a simple example, i.e., single-cavity case ($N=1$). The discussion will be structured around three main aspects: {\it I. Physical interpretation; II. Analytic solutions; III. Numerical simulations and verification.}

{\it I: Physical interpretation.--} Here, we provide detailed descriptions of two distinct configurations of the waveguide-coupled cavity, as depicted in Fig.\,\ref{Fig_Sup7}. The first configuration, referred to as the direct-coupled cavity (DCC), features a singular multi-photon scattering pathway, i.e., multiple photons are first absorbed and then re-emitted by the cavity. The second configuration, termed the side-coupled cavity (SCC), allows for more possible scattering paths. Owing to the unique properties of the SCC, incident photons can freely propagate to the output port. Consequently, the two-photon scattering paths in the SCC can be categorized into three scenarios: two-photon free propagation, two-photon absorption and re-emission, and one photon freely propagating while the other is absorbed and re-emitted. More importantly, we also provide the probability amplitude of each possible scattering pathway by employing the $S$-Matrix theory, i.e.,
\begin{align}\label{VI50}
	\mathcal{I}_1=e^{-i\omega_d t},\quad P^1_1=O^{r}_{0,1}\mathcal{K}_{\omega_d}^{-1}(1)O_{0,1}^{r\dagger}e^{-i\omega_d t},\quad P^1_2=O^{r}_{0,1}O^{r}_{1,2}\mathcal{K}_{2\omega_d}^{-1}(2)O_{1,2}^{r\dagger}\mathcal{K}_{\omega_d}^{-1}(1)O_{0,1}^{r\dagger}e^{-2i\omega_d t},
\end{align}
where
\begin{align}\label{VI51}
	O^{r}_{0,1}=\sqrt{\kappa_r}\mqty[1&0],\quad  O^{r}_{1,2}=\sqrt{\kappa_r}\mqty[\sqrt{2}&0\\0&1],\quad  \mathcal{K}_{n\omega_d}(n)=-i\mqty[n(\Delta-i\kappa/2)&\text{g}\sqrt{n}\\\text{g}\sqrt{n}&n(\Delta-i\kappa/2)+i(\kappa-\gamma_e)/2].
\end{align}
Here, we have assumed a coupling efficiency of 1 and that the atomic transition frequency is resonant with the cavity's free frequency, i.e., $\kappa_e=0$ and $\Delta=\omega_c-\omega_d=\omega_e-\omega_d$.

{\it II: Analytic solutions.--} Subsequently, according to Eqs.~(\ref{VI50}-\ref{VI51}), in the weak driving approximation, the second-order equal-time correlation function of transmitted light for the two configurations is computed as
\begin{align}\label{VI52}
	g_{rr,\text{DCC}}^{(2)}(0)=\frac{\abs{P_2^1}^2}{\abs{P_1^1}^4}=\frac{\big|{P}_2^1(0)\big|^2}{\big|{P}_1^1(0)\big|^4}, \quad g_{rr,\text{SCC}}^{(2)}(0)=\frac{\abs{\mathcal{I}_2+2P_1^1\mathcal{I}_1+P_2^1}^2}{\abs{\mathcal{I}_1+P_1^1}^4}=\frac{\big|1+2{P}_1^1(0)+{P}_2^1(0)\big|^2}{\big|1+{P}_1^1(0)\big|^4}.
\end{align}
It is evident that for the DCC and SCC configurations, the transmitted light exhibits bunched and antibunched behaviors, respectively, when $|{P}^1_1(0)|\ll \min\{1, |{P}_2^1(0)|\}$ and ${P}_2^1(0)\approx -1$. To ensure that both conditions can be satisfied simultaneously, by plugging Eq.~(\ref{VI51}) into Eq.~(\ref{VI50}), the time-independent single- and two-photon probability amplitudes at zero detuning ($\Delta=0$) can be written as\vspace{-5pt}
\begin{align}\label{VI53}
	{P}_1^1(0)=\frac{\kappa_r}{\kappa}\frac{-2\gamma_e/\kappa}{4(\mathrm{g}/\kappa)^2+\gamma_e/\kappa}, \quad {P}_2^1(0)=\left(\frac{\kappa_r}{\kappa}\right)^2\frac{-16(\mathrm{g}/\kappa)^2+(\gamma_e/\kappa+1)\gamma_e/\kappa}{[4(\mathrm{g}/\kappa)^2+\gamma_e/\kappa][4(\mathrm{g}/\kappa)^2+\gamma_e/\kappa+1]}.
\end{align}
When $\gamma_e\ll\mathrm{g}$, Eq.~(\ref{VI53}) can be further simplified as\vspace{-5pt}
\begin{align}
	|{P}_1^1(0)|\ll 1,\quad {P}_2^1(0)\approx -\frac{1}{(1+\alpha)^2}\frac{4}{4(\mathrm{g}/\kappa)^2+1}.
\end{align} 
Thus, for an appropriate coupling strength $\mathrm{g}$, ${P}_2^1(0)$ can always be adjusted to approach minus one by modulating the chirality $\alpha$. This results in markedly distinct statistical properties (i.e., photons bunching and antibunching) of the transmitted light in the two configurations. In fact, these differences mainly stem from the difference in the number of two-photon scattering paths in the two different configurations. As shown by the incident two-photon scattering paths in Fig.\,\ref{Fig_Sup7}, the two additional scattering paths in the SCC configuration, compared to the DCC configuration, can form destructive interference with the remaining pathway, and the conditions for this interference to occur are as discussed above.

\begin{figure}
	\includegraphics[width=14cm]{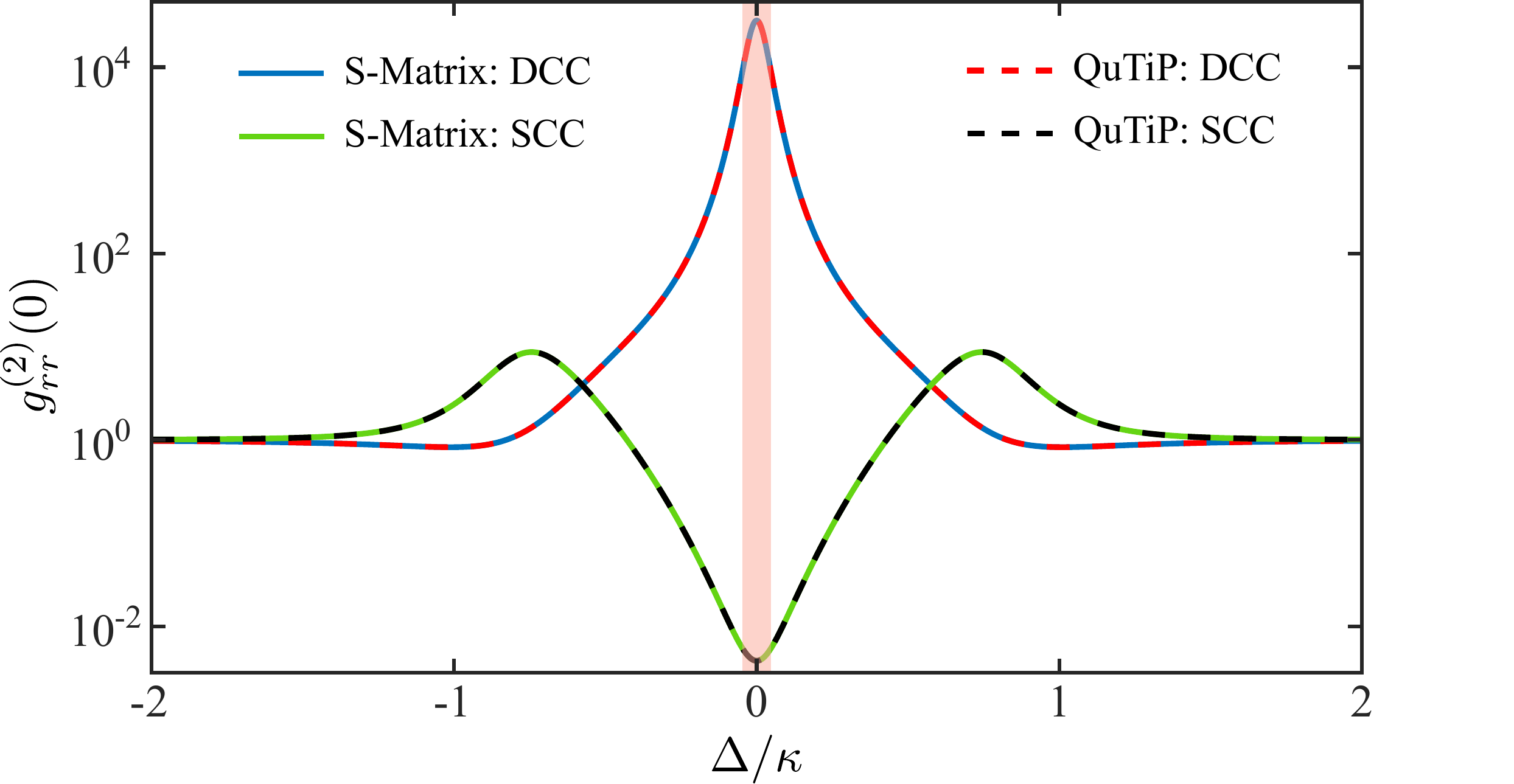}\\
	\caption{The second-order equal-time correlation function $g_{rr}^{(2)}(0)$ versus the detuning $\Delta/\kappa$. The solid and dotted lines represent the result obtained by using the $S$-Matrix approach and the master equation (done using QuTiP), respectively. Here, the system parameters are given by $\text{g}/\kappa=0.8$ and $\gamma_e/\kappa=0.1$. The driving strength $\Omega/\kappa=10^{-3}$ for QuTiP simulation, and the chirality $\alpha=0.05$ for the SCC configuration.
	}\label{Fig_Sup8}
\end{figure}

{\it III: Numerical simulations and verification.--} As depicted in Fig.\,\ref{Fig_Sup8}, the statistical properties of the transmitted light in the two configurations exhibit entirely opposite behaviors: strong photons bunching ($g_{rr,\text{DCC}}^{(2)}(0)\gg1$) in DCC configuration and strong photons antibunching ($g_{rr, \text{SCC}}^{(2)}(0)\ll 1$) in SCC configuration when the system parameter, specifically the detuning, approaches zero. Additionally, we also see in Fig.\,\,\ref{Fig_Sup8} that the QuTiP simulations agree perfectly with the $S$-Matrix based simulations, thereby ensuring the correctness of our work. Moreover, even in the multiple cavities case, our results obtained using the $S$-Matrix method remain accurate, as detailed in Section \ref{X}.

Finally, we conclude that the statistical properties of the transmitted light differ significantly between the two configurations, even within the same model. In the DCC configuration, the behavior of the transmitted light depends solely on system parameters such as detuning and coupling strength. Conversely, in the SCC configuration, the behavior of the transmitted light is influenced not only by these parameters but also by chirality. Consequently, the transmitted light described in Refs.\,\cite{faraon_coherent_2008, birnbaum_photon_2005} is based on the DCC configuration and thus exhibits bunched behavior at zero detuning, as shown by the blue line within the pink region in Fig.\,\ref{Fig_Sup8}. However, the transmitted light in the main text is based on the SCC configuration and thus shows anti-bunched behavior at zero detuning, as indicated by the green line within the pink region in Fig.\,\ref{Fig_Sup8}. 

\section{Waveguide Side-Coupled to a Two-Level Atom System and a Cavity-Atom System}
\begin{figure}
	\includegraphics[width=17cm]{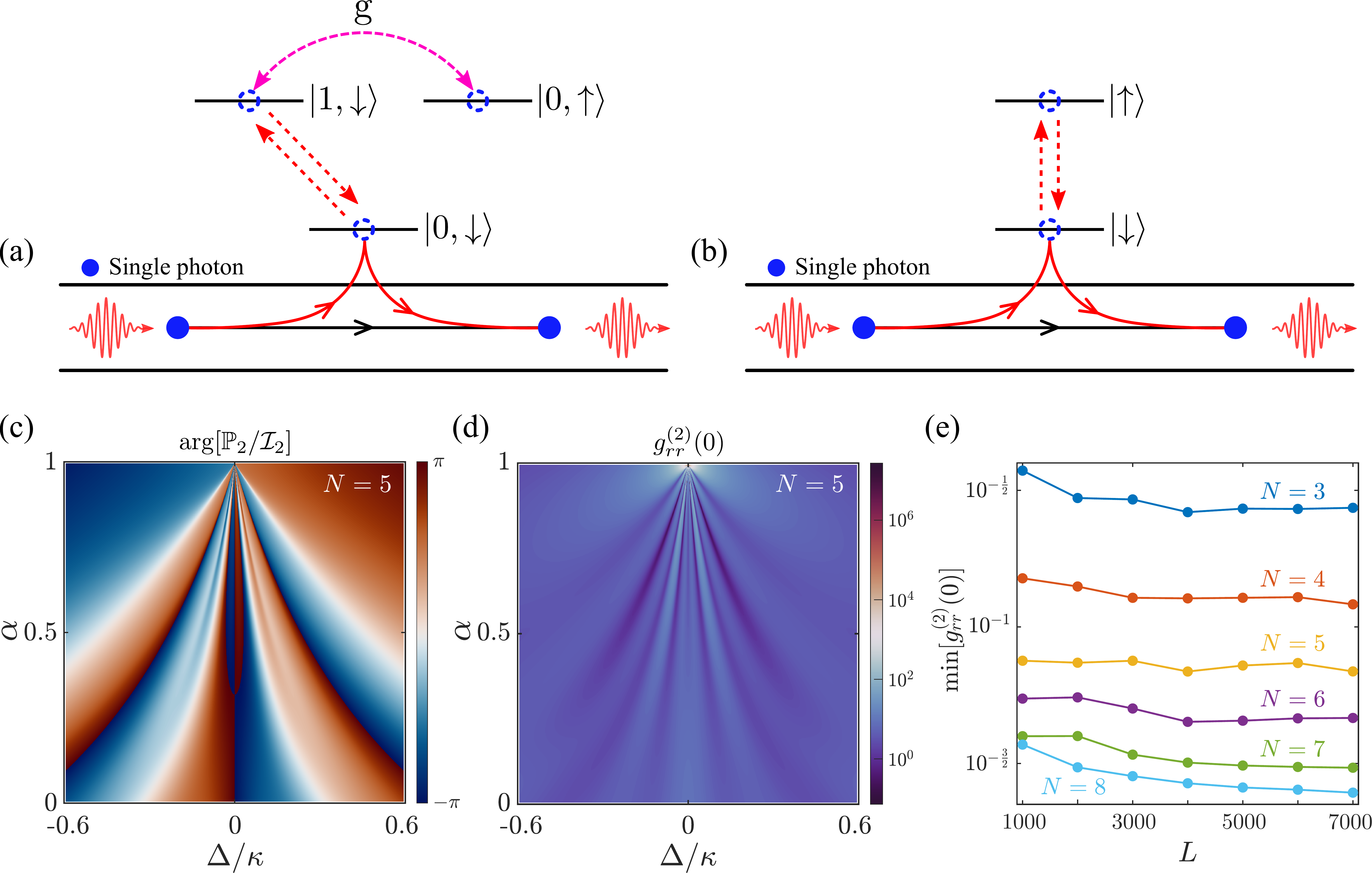}\\
	\caption{Single-photon scattering process in the system with (a) a cavity containing a two-level atom, coupled to a waveguide, and (b) a two-level atom coupled to a waveguide. Here, red and black lines represent the scattering paths, pink dashed in (a) represent transition paths, and the first and last indexes in $\ket{n,\downarrow(\uparrow)}$ denote $n$-photon Fock state of the cavity and ground (excited) state of the two-level atom, respectively. The complex argument of the time-independent two-photon total probability amplitude $\arg[\mathbb{P}_2/\mathcal{I}_2]$ in (c) and the equal-time second-order quantum correlation $g_{rr}^{(2)}(0)$ in (d) versus the chirality $\alpha$ and the detuning $\Delta/\kappa$, for the number of atoms $N=5$. (e) The minimum second-order quantum correlation as a function of sampling frequency $L$, for the system with $N=3,4,5,6,7$, and $8$ atoms. All results in (c-e) are obtained from the system with $N$ two-level atoms chirally coupled to a waveguide.
	}\label{Fig_Sup9}
\end{figure}
In this section, we will elucidate the underlying physical mechanisms using a single-cavity side-coupled waveguide system as a simple example, highlighting its subtle differences from the unconventional PB mechanism. Additionally, we will discuss in detail the scenario of a single two-level atom side coupled to a waveguide and extend our discussion to multi-atom systems.

Similar to Eq.~(1) in the main text, the total Hamiltonian is given by $H_{\text{tot},\ell }=H_{\text{sys},\ell}+H_{\text{wg}}+H_{int,\ell}$, and
\begin{align}
	H_{\text{sys},1}&=\sum_{j=1}^{N}[\omega_c a_j^\dagger a_j^{}+\omega_e \sigma_j^\dagger\sigma_j^{}+\text{g}(a_j^\dagger\sigma_j^{}+\sigma_j^\dagger a_j^{})],\quad
	H_{\text{sys},2}=\sum_{j=1}^{N}\omega_e \sigma_j^\dagger\sigma_j^{},\quad H_{\text{wg}}=\sum_{\mu=l,r}{\int\dd\omega\ \omega b^\dagger_\mu(\omega)b^{}_\mu(\omega)},\\
	H_{\text{int}, 1} &=\sum_{j=1}^N\sum_{\mu=l,r}\sqrt{\frac{\kappa_\mu}{2\pi}}\int\dd\omega[b_\mu^{}(\omega)a^{\dagger}_je^{i\omega x_j/v_\mu}+\text{h.c.}],\quad H_{\text{int}, 2} =\sum_{j=1}^N\sum_{\mu=l,r}\sqrt{\frac{\kappa_\mu}{2\pi}}\int\dd\omega[b_\mu^{}(\omega)\sigma^{\dagger}_je^{i\omega x_j/v_\mu}+\text{h.c.}],
\end{align}
where the subscript $\ell=1(2)$ represents $N$ cavities (atoms) side-coupled to a waveguide. Following the steps of Eqs.\,(\ref{eq1}-\ref{eq23}), we can derive two corresponding effective Hamiltonian, given by
\begin{align}
	H_{\text{eff},1}&=H_{\text{sys},1}-i\sum_{j=1}^N\frac{\kappa_r+\kappa_l}{2}a_{j}^\dagger a_{j}^{}-i\sum_{j>k}^Ne^{i\phi(j-k)}[\kappa_ra_{j}^\dagger a_{k}^{}+\kappa_la_{k}^\dagger a_{j}^{}],\\
	H_{\text{eff},2}&=H_{\text{sys},2}-i\sum_{j=1}^N\frac{\kappa_r+\kappa_l}{2}\sigma_{j}^\dagger \sigma_{j}^{}-i\sum_{j>k}^Ne^{i\phi(j-k)}[\kappa_r\sigma_{j}^\dagger \sigma_{k}^{}+\kappa_l\sigma_{k}^\dagger \sigma_{j}^{}].
\end{align}
Note that in the following discussion regarding the physical mechanism, we only consider the case of $N=1$ and focus on a single-photon scattering process within the waveguide. 

By comparing Figs.\,\ref{Fig_Sup9}(a) and \ref{Fig_Sup9}(b), we observe that the system with a cavity containing a two-level atom, coupled to the waveguide, exhibits significant differences in these types of scattering paths, i.e., red dashed in Fig.\,\ref{Fig_Sup9}(a), where photons are first absorbed by a cavity and then re-emitted into the waveguide. Specifically, photons absorbed by the cavity can undergo a series of possible transition paths, i.e., pink dashed in Fig.\,\ref{Fig_Sup9}(a). For example, a photon is first absorbed by the cavity, then, through the coupling between the atom and the cavity, it transitions back and forth between the atom and the cavity before being finally re-emitted into the waveguide, as shown in Fig.\,\ref{Fig_Sup9}(a). Such transition paths do not exist for the system of only two-level atoms coupled to the waveguide, as shown in Fig.\,\ref{Fig_Sup9}(b). Although we have analyzed only the single-photon scattering process here, a similar analysis can be applied to multi-photon scattering, leading to the same conclusions. Here, the transition paths are exactly the interference paths involved in unconventional PB. In conclusion, compared to others studies \cite{PhysRevLett.121.043601, PhysRevLett.121.043602, PhysRevLett.127.240402, prasad_correlating_2020, PhysRevLett.121.143601}, where PB is solely attributed to the destructive interference of either scattering paths or transition paths, the underlying physical mechanism of the main text involves the superposition of both two types of paths.

Additionally, due to the different energy level structures, as shown in Figs.\,\ref{Fig_Sup9}(a) and \ref{Fig_Sup9}(b), we conclude that not all systems can produce the results presented in the main text. For example, for the system with $N$ two-level atoms chirally coupled to the waveguide, we do not observe phase singularity, as shown in Fig.\,\ref{Fig_Sup9}(c), indicating that the second-order quantum correlation $g_{rr}^{(2)}(0)$ has no analytic zeros, as illustrated by Fig.\,\ref{Fig_Sup9}(d). Note that the absence of our paper's results in this system is not due to having too few atoms or sparsely distributed data points. As shown in Fig.\,\ref{Fig_Sup9}(e), on the one hand, as the data points become denser (i.e., increasing $L$), we observe that the minimum value of the equal-time second-order quantum correlation remains almost unchanged. On the other hand, as the number of atoms increases, although the minimum value of the correlation function gradually decreases, the rate of decrease becomes smaller. Even for a lager number of atoms, it is hard to reach the value required for near-perfect PB phenomenon.

\section{Proof of Eq.(6)}
To support Eq.(9) in the main text, in this section, we provide the detailed derivation for the single-photon probability amplitude. For definiteness and without loss of generality, we assume that each atom has a frequency $\omega_e$ and dissipation $\gamma_e$, each cavity has a frequency $\omega_c$ and external dissipation $\kappa_c$, and the incident light has a frequency $\omega_d$. Note that we only consider a mirror configuration in this section. According to Sec.~\ref{IIB} and the effective Hamiltonian~(\ref{eq47}), in the single-excitation number subspace, the effective Hamiltonian is given by
\begin{align}\label{V48}
	H_{N,\text{eff}}^{(1)}=\mqty[H_{N-1,\text{eff}}^{(1)} & B_{N-1,l}^{T}\\ B_{N-1,r} & H_{1,\text{eff}}^{(1)}],\quad H_{1,\text{eff}}^{(1)}=\mqty[\omega_c-i(\kappa+\kappa_c)/2&\mathrm{g}\\ \mathrm{g} &\omega_e-i\gamma_e/2].
\end{align}
For the sake of simplicity, we define 
\begin{align}\label{V49}
	x_N=\left[\begin{matrix}
		X_1&X_1&\cdots&X_1
	\end{matrix}\right]_{1\times N}.
\end{align}
According to Eqs.~(\ref{eq35}, \ref{eq38}), it is easy to see that $B_{N, \mu}=-i\kappa_\mu x_N\otimes X_1^T $ and $O^r_{0,1}=\sqrt{\kappa_r}x_N$. Similarly, we have
\begin{align}\label{V50}
	\mathcal{H}_N^{-1}=[H_{N,\text{eff}}^{(1)}-\omega_d]^{-1}=\mqty[\mathcal{H}_{N-1}^{-1}+ \mathcal{H}_{N-1}^{-1} B_{N-1,l}^T \tilde{Q}^{-1}B_{N-1,r} \mathcal{H}_{N-1}^{-1} && -\mathcal{H}_{N-1}^{-1} B_{N-1,l}^T \tilde{Q}^{-1} \\ -\tilde{Q}^{-1}B_{N-1,r} \mathcal{H}_{N-1}^{-1}&& \tilde{Q}^{-1}],
\end{align}
where $\tilde{Q}=\mathcal{H}_1 - B_{N-1,r} \mathcal{H}_{N-1}^{-1} B_{N-1,l}^{T}$. Then, by defining $y_N=x_N\mathcal{H}_N^{-1}x_N^T$, we have
\begin{align}
	y_N&=\mqty[x_{N-1} & x_1] \mqty[\mathcal{H}_{N-1}^{-1}+ \mathcal{H}_{N-1}^{-1} B_{N-1,l}^T \tilde{Q}^{-1}B_{N-1,r} \mathcal{H}_{N-1}^{-1} && -\mathcal{H}_{N-1}^{-1} B_{N-1,l}^T \tilde{Q}^{-1} \\ -\tilde{Q}^{-1}B_{N-1,r} \mathcal{H}_{N-1}^{-1}&& \tilde{Q}^{-1}] \mqty[x_{N-1}^T \\ x_1^T]\nonumber\\
	&=y_{N-1}-\kappa_l\kappa_r \mqty[y_{N-1}&0 ] \tilde{Q}^{-1} \mqty[y_{N-1}\\0 ] +i\kappa_r x_1\tilde{Q}^{-1}\mqty[y_{N-1}\\0] +i\kappa_l\mqty[y_{N-1}&0] \tilde{Q}^{-1}x_{1}^T + x_1\tilde{Q}^{-1}x_1^T,\label{V51}
\end{align}
and 
\begin{align}\label{V52}
	\tilde{Q}=\mathcal{H}_1+\kappa_l\kappa_r\mqty[y_{N-1}&0\\0&0]=\mqty[\tilde{\Delta}_c+\kappa_l\kappa_r y_{N-1}&&\mathrm{g}\\ \mathrm{g}&& \tilde{\Delta}_e ],
\end{align}
where $\tilde{\Delta}_c=\omega_c-\omega_d-i(\kappa+\kappa_c)/2$ and $\tilde{\Delta}_e=\omega_e-\omega_d-i\gamma_e/2$. Therefore, Eq.~(\ref{V51}) can be further simplified as
\begin{align}
	y_{N}&=y_{N-1}-\frac{\kappa_l\kappa_r\tilde{\Delta}_e}{\text{det}[\tilde{Q}]}y_{N-1}^2+\frac{i\kappa_r\tilde{\Delta}_e}{\text{det}[\tilde{Q}]}y_{N-1}+\frac{i\kappa_l\tilde{\Delta}_e}{\text{det}[\tilde{Q}]}y_{N-1}+\frac{\tilde{\Delta}_e}{\text{det}[\tilde{Q}]}y_{N-1}=y_{N-1}+\frac{\tilde{\Delta}_e(1-\kappa_l\kappa_r y_{N-1}^2+i\kappa y_{N-1})}{\text{det}[\tilde{Q}]}\nonumber\\
	&=y_{N-1}+\frac{\tilde{\Delta}_e(1-\kappa_l\kappa_r y_{N-1}^2+i\kappa y_{N-1})}{\tilde{\Delta}_c\tilde{\Delta}_e-\mathrm{g}^2+\kappa_l\kappa_r\tilde{\Delta}_e y_{N-1}}=y_{N-1}+\frac{1-\xi_1 y_{N-1}^2+i\kappa y_{N-1}}{\xi_2+\xi_1y_{N-1}},\label{V53}
\end{align}
where $\xi_1=\kappa_l\kappa_r$ and $\xi_2=(\tilde{\Delta}_c\tilde{\Delta}_e-\mathrm{g}^2)/\tilde{\Delta}_e$. According to $P^{N}_1(0)=O^{r}_{0,1}\mathcal{K}_{\omega_d}^{-1}(1)O_{0,1}^{r\dagger} =i\kappa_ry_N$, by solving the recurrence relation~(\ref{V53}), the time-independent single-photon probability amplitude is 
\begin{align}\label{V54}
	P^{N}_1(0)=2i\kappa_r\left[\sqrt{4\xi_1-\kappa^2}-i\kappa+\frac{2\sqrt{4\xi_1-\kappa^2}}{-1+(\textbf{r}e^{i\theta})^N}\right]^{-1}=2\left[ \abs{1-\alpha}\frac{(\textbf{r}e^{i\theta})^N+1}{(\textbf{r}e^{i\theta})^N-1}  -(1+\alpha)\right]^{-1},
\end{align}
where
\begin{align}
	\textbf{r}=\abs{\frac{2\xi_2+i\kappa+\sqrt{4\xi_1-\kappa^2}}{2\xi_2+i\kappa-\sqrt{4\xi_1-\kappa^2}}},\quad \theta=\arg\left(\frac{2\xi_2+i\kappa+\sqrt{4\xi_1-\kappa^2}}{2\xi_2+i\kappa-\sqrt{4\xi_1-\kappa^2}}\right).
\end{align}
We note that $\textbf{r}\le 1$ for any parameters, and $\textbf{r}=1$ only when $\gamma_e=\kappa_c=0$. On the one hand, for the parameters chosen in the main text, i.e., $\Delta=\omega_c-\omega_d=\omega_e-\omega_d$, $\gamma_e=\kappa_c=0$, and $\alpha\in[0, 1]$, Eq.~(\ref{V54}) can be further simplified as
\begin{align}
	P_{1}^N(0)=P_1^Ne^{i\omega_dt}=-\frac{1-\exp(iN\theta)}{1-\alpha \exp(iN\theta)},\ \ \text{with}\ \  \theta=2\arctan\left[\frac{\Delta(1-\alpha)\kappa}{2(1+\alpha)(\Delta^2-\mathrm{g}^2)}\right],
\end{align}
corresponding to Eq.(9) in the main text. On the other hand, for cases with external loss channels, we find that the single-photon transmission (being scattered into the right-side port of the waveguide) and reflection (being scattered into the left-side port of the waveguide) only depend on the chirality $\alpha$ due to $\textbf{r}^N\to0$ when $N\gg1$. To determine the probability of a single photon not being scattered into external loss channels, i.e., the survival rate (the probability of being scattering into the waveguide), we need to calculate the sum of single-photon transmission and reflection. The result is 
\begin{align}
	T+R= \abs{1 + O^{r}_{0,1}\mathcal{K}_{\omega_d}^{-1}(1)O_{0,1}^{r\dagger}}^2 + \abs{O^{l}_{0,1}\mathcal{K}_{\omega_d}^{-1}(1)O_{0,1}^{r\dagger}}^2=\abs{1+P_1^{N}(0)}^2+\abs{P_1^N(0)\sqrt{\kappa_l/\kappa_r}}^2,
\end{align}
according to Eq.\,(\ref{V54}), and we further have 
\begin{align}
	\lim\limits_{N\to\infty}(T+R)=\abs{1-\frac{2}{\abs{1-\alpha}+1+\alpha}}^2+\abs{-\frac{2\sqrt{\alpha}}{\abs{1-\alpha}+1+\alpha}}^2=\begin{cases}
		\alpha&\quad\alpha\le1\\
		1-\alpha^{-1}+\alpha^{-2}&\quad\alpha>1
	\end{cases}.
\end{align}
We note that the survival rate is equal to zero only when $\alpha=0$, and it exhibits an exponential decrease with the number of cavities in the fully chiral condition, i.e., $\ln(T+R)=2N\ln\abs{\textbf{r}}$, as shown by the blue line in Fig.\,\ref{Fig_Sup5}(b), while in the weakly chiral condition, the survival rate can reach a high value when $N\gg1$, as shown by the yellow line in Fig.\,\ref{Fig_Sup5}(b).
\vspace{-10pt}
\section{Validating the Scattering Matrix Calculation for the Single Cavity Case}
In this section, we provide a detailed derivation to construct the connection between the $S$-Matrix method and the Lindblad master equation, and we also numerically verify that the equal-time two-photon-correlation expressions derived in the main text matches with a master-equation based simulation of the single cavity-atom system. 

Firstly, according to Eq.~(1) in the main text, the total Hamiltonian for $N=1$ reads as 
\begin{align}
	H_{\rm tot}=H_{\rm sys} + H_{\rm wg} + H_{\rm int},
\end{align}
where $H_{\rm sys}=\omega_c a^\dagger a + \omega_e\sigma^\dagger\sigma+\text{g}\left(a^\dagger\sigma + \sigma^\dagger a \right)$ and $H_{\rm int}=\sum_{\mu=l,r}\sqrt{\frac{\kappa_\mu}{2\pi}}\int\dd{\omega}[b_\mu^\dagger(\omega)a+a^\dagger b_{\mu}(\omega)]$. Before constructing the bridge, we need to understand some concepts of the $S$-Matrix theory. Specifically, the scattering matrix satisfies $S=\Omega_-^\dagger\Omega_+^{}$, where $\Omega_{\pm}=\exp(iH_{\rm tot}t_{\pm})\exp(-iH_{\rm wg}t_{\pm})$ with $t_\pm\to\mp\infty$, called the Møller wave operators. We can directly obtain the input-output operators through the Møller wave operators, i.e.,
\begin{align}
	b_{\mu,\text{in}}(t)&\equiv\Omega_+ b_{\mu}(t)\Omega_+^\dagger=e^{iH_{\rm tot}t_+}e^{-iH_{\rm wg}t_+}b_{\mu}(t)e^{iH_{\rm wg}t_+}e^{-iH_{\rm tot}t_+},\\
	b_{\mu,\text{out}}(t)&\equiv\Omega_-b_{\mu}(t)\Omega_-^\dagger=e^{iH_{\rm tot}t_-}e^{-iH_{\rm wg}t_-}b_{\mu}(t)e^{iH_{\rm wg}t_-}e^{-iH_{\rm tot}t_-},
\end{align}
and the input-output relations satisfy $b_{\mu,\text{out}}(t)=b_{\mu,\text{in}}(t)-iA_{\mu}(t)=b_{\mu,\text{in}}(t)-i\sqrt{\kappa_\mu}a(t)$ corresponding to Eq.~(\ref{eq9}). Besides, the quantum causality condition about $A_{\mu}(t)$ and $A_{\mu}^\dagger(t)$ at equal time is $[A_{\mu}(t), b_{\nu,\text{in}}(t)]=[A_{\mu}^\dagger(t),b_{\nu,\text{in}}(t)]=0$. The detailed proofs of these concepts can be found in \cite{PhysRevA.108.053703}. Based on these conclusions, we have
\begin{align}
	\langle \psi_{\text{out}}| b_r^{\dagger n}(t)b_r^{n}(t) |\psi_{\text{out}}\rangle&= \langle\psi_{\text{in}}| S^\dagger \Omega_-^\dagger [\Omega_- b_r^\dagger(t)\Omega_-^\dagger]^n[\Omega_-b_r(t)\Omega_-^\dagger]^n\Omega_-S |\psi_{\text{in}}\rangle =\langle \psi_{\text{in}}| \Omega_+^\dagger b_{r,\text{out}}^{\dagger n}(t) b_{r,\text{out}}^n(t)\Omega_+ |\psi_{\text{in}}\rangle\nonumber\\
	&=\langle \psi_{\text{in}}| \Omega_+^\dagger [b_{r,\text{in}}^\dagger(t)+iA_{r}^\dagger(t)]^n[b_{r,\text{in}}(t)-iA_{r}(t)]^n \Omega_+ |\psi_{\text{in}}\rangle.\label{IX73}
\end{align}
According to these definitions of $b_{r,\text{in}}(t)$, $\Omega_+$, and $|\psi_{\text{in}}\rangle$, we have
\begin{align}
	b_{r,\text{in}}(t)\Omega_+|\psi_{\text{in}}\rangle&=\frac{1}{\sqrt{2\pi}}\int_{-\infty}^{+\infty}e^{iH_{\rm tot}t_+}e^{-iH_{\rm wg}t_+} b_r(\omega)e^{iH_{\rm wg}t_+}e^{-iH_{\rm tot}t_+}e^{-i\omega t}\dd{\omega} \left[e^{iH_{\rm tot}t_+}e^{-iH_{\rm wg}t_+} \right] |\eta\rangle_{r}^{\omega_d}\otimes \ket{0}_l\otimes\ket{0}_s\nonumber\\
	&=\frac{1}{\sqrt{2\pi}}\int_{-\infty}^{+\infty}e^{iH_{\rm tot}t_+}e^{-iH_{\rm wg}t_+} b_r(\omega)e^{-i\omega t}|\eta\rangle_{r}^{\omega_d}\otimes \ket{0}_l\otimes\ket{0}_s\dd{\omega}\nonumber\\
	&=\frac{\eta}{\sqrt{2\pi}}e^{iH_{\rm tot}t_+}e^{-iH_{\rm wg}t_+} e^{-i\omega_d t}|\eta\rangle_{r}^{\omega_d}\otimes \ket{0}_l\otimes\ket{0}_s=\tilde{\eta} e^{-i\omega_d t}\Omega_+|\psi_{\text{in}}\rangle,\label{IX74}
\end{align}
where $\tilde{\eta}=\eta/\customsqrt{2\pi}$ and $|\psi_{\text{in}}\rangle=\mathcal{N}\sum_{k=0}^\infty[\eta^kb_{r}^{\dagger k}(\omega_d)/k!]\ket{0}_r\otimes\ket{0}_l\otimes\ket{0}_s=\ket{\eta}_{r}^{\omega_d}\otimes\ket{0}_l\otimes\ket{0}_s$ corresponding to a coherent state with amplitude $\eta$ and frequency $\omega_d$ entering into the waveguide from the left-side port. To further simplify Eq.~(\ref{IX73}), one needs to compute $[b_{r,\text{in}}(t)-iA_{r}(t)]^n \Omega_+ |\psi_{\text{in}}\rangle$. According to Eq.~(\ref{IX74}) and the quantum causality condition, we have
\begin{align}
	[b_{r,\text{in}}(t)-iA_{r}(t)]^n \Omega_+ |\psi_{\text{in}}\rangle&=[b_{r,\text{in}}(t)-iA_{r}(t)]^{n-1}[\tilde{\eta} e^{-i\omega_d t}-iA_{r}(t)]\Omega_+|\psi_{\text{in}}\rangle \nonumber\\
	&=[\tilde{\eta} e^{-i\omega_d t}-iA_{r}(t)][b_{r,\text{in}}(t)-iA_{r}(t)]^{n-1}\Omega_+|\psi_{\text{in}}\rangle \nonumber\\
	&=[\tilde{\eta} e^{-i\omega_d t}-iA_{r}(t)]^{n}\Omega_+|\psi_{\text{in}}\rangle=e^{-in\omega_d t}[\tilde{\eta} -iA_{r}(t)e^{i\omega_d t}]^{n}\Omega_+|\psi_{\text{in}}\rangle,
\end{align} 
and Eq.~(\ref{IX73}) can be further computed as
\begin{align}
	G^{(n)}_{rr}(0)&=\langle \psi_{\text{out}}| b_r^{\dagger n}(t)b_r^{n}(t) |\psi_{\text{out}}\rangle=\langle\psi_{\text{in}}|\Omega_+^\dagger[\tilde{\eta}^*+iA_{r}^\dagger(t)e^{-i\omega_d t}]^n[\tilde{\eta} -iA_{r}(t)e^{i\omega_d t}]^{n}\Omega_+|\psi_{\text{in}}\rangle\nonumber\\
	&=\Tr{[\tilde{\eta}^*+iA_{r}^\dagger(t)e^{-i\omega_d t}]^n[\tilde{\eta} -iA_{r}(t)e^{i\omega_d t}]^{n}\Omega_+|\psi_{\text{in}}\rangle\langle\psi_{\text{in}}|\Omega_+^\dagger}\nonumber\\
	&=\Tr{ e^{iH_{\rm tot}t}[\tilde{\eta}^*+iA_{r}^\dagger e^{-i\omega_d t}]^n[\tilde{\eta} -iA_{r}e^{i\omega_d t}]^{n}e^{-iH_{\rm tot}t}  e^{iH_{\rm tot}t_+}e^{-iH_{\rm wg}t_+}  |\psi_{\text{in}}\rangle\langle\psi_{\text{in}}|e^{iH_{\rm wg}t_+}e^{-iH_{\rm tot}t_+}}\nonumber\\
	&=\Tr{[\tilde{\eta}^*+iA_{r}^\dagger e^{-i\omega_d t}]^n[\tilde{\eta} -iA_{r}e^{i\omega_d t}]^{n}e^{-iH_{\rm tot}(t-t_+)}  \rho(0)e^{iH_{\rm tot}(t-t_+)}}\nonumber\\
	&=\Tr_{\rm S}\left\{[\tilde{\eta}^*+iA_{r}^\dagger e^{-i\omega_d t}]^n[\tilde{\eta} -iA_{r}e^{i\omega_d t}]^{n} \rho_s(\tau)\right\}=\Tr_{\rm S}\left\{[\tilde{\eta}^*+iA_{r}^\dagger]^n[\tilde{\eta} -iA_{r}]^{n}[\mathcal{U}(t)\rho_s(\tau)\mathcal{U}^\dagger(t)]\right\},\label{IX76}
\end{align}
where $\rho_{s}(\tau)=\Tr_{\rm B}[U(\tau,0)\rho(0)U^\dagger(\tau,0)]$, $U(\tau,0)=\exp(-iH_{\rm tot}\tau)$, $\rho(0)=\exp(-iH_{\rm wg}t_+)|\psi_{\text{in}}\rangle\langle\psi_{\text{in}}|\exp(iH_{\rm wg}t_+)$, $\Tr_{{\rm S}({\rm B})}$ represents partial trace for systems (baths), and $\tau=t-t_+$. Note that in Eq.~(\ref{IX76}), the rotating-frame transformation satisfies $\mathcal{U}^\dagger(t) A_r\mathcal{U}^{}(t)=A_re^{i\omega_d t}$ and $\mathcal{U}^\dagger(t)\mathcal{U}(t)=\mathbb{I}$. Subsequently, we introduce a time-dependent displacement operator
\begin{align}
	\mathcal{D}_t\{\eta(\omega)\}=\exp(\int_{-\infty}^{+\infty}\dd{\omega}\left[\eta(\omega)e^{-i\omega t}b_{r}^\dagger(\omega)-\eta^*(\omega)e^{i\omega t}b_r(\omega)\right]),
\end{align}
where $\eta(\omega)=\eta\delta(\omega-\omega_d)$. Thus, the free evolution of input coherent state can be rewritten as $\exp(-iH_{\rm B}t_+)|\psi_{\text{in}}\rangle=\mathcal{D}_0\{\eta(\omega)\exp(-i\omega t_+)\}$, and the reduced density matrix $\rho_s(\tau)$ can be simplified as
\begin{align}
	\rho_s(\tau)&=\Tr_{\rm B}\left[\mathcal{D}_\tau^\dagger\{\eta(\omega)e^{-i\omega t_+}\}  U(\tau, 0) \mathcal{D}_0\{\eta(\omega)e^{-i\omega t_+}\} \tilde{\rho}(0) \mathcal{D}_0^\dagger\{\eta(\omega)e^{-i\omega t_+}\} U^\dagger(\tau, 0) \mathcal{D}_\tau\{\eta(\omega)e^{-i\omega t_+}\}  \right]\nonumber\\
	&=\Tr_{\rm B}\left[ \tilde{U}(\tau, 0)\tilde{\rho}(0)\tilde{U}^\dagger(\tau, 0)\right]=\tilde{\rho}_s(\tau),
\end{align}
where $\tilde{\rho}(0)=|0\rangle_r\langle0|\otimes|0\rangle_l\langle0|\otimes|0\rangle_s\langle0|$ and $\tilde{U}(\tau, 0)=\mathcal{D}_\tau^\dagger\{\eta(\omega)e^{-i\omega t_+}\} U(\tau, 0) \mathcal{D}_0\{\eta(\omega)e^{-i\omega t_+}\}$. And beyond that, according to the Mollow transformation \cite{PhysRevA.12.1919}, the new evolution operator $\tilde{U}(\tau,0)$ satisfies
\begin{align}\label{IX79}
	i\dv{\ }{\tau}\tilde{U}(\tau, 0)=\tilde{H}(\tau)\tilde{U}(\tau, 0),\ \text{with},\ \tilde{H}(\tau)=H_{\rm tot}+H_{\rm d}(\tau+t_+),
\end{align}
where 
\begin{align}
	H_{\rm d}(\tau+t_+)=H_{\rm d}(t)=\tilde{\eta} A_r^\dagger e^{-i\omega_d t} + \tilde{\eta}^* A_r e^{i\omega_d t}=\left(\Omega a^\dagger e^{-i\omega_dt} + \text{H.c.}\right)
\end{align}
describing the external coherent drive with strength $\Omega=\tilde{\eta}\sqrt{\kappa_r}$. Finally, based on Eq.~(\ref{IX79}), following the standard procedures \cite{Pathria1996, lidar_lecture_2020} to trace out the baths degrees of freedom and applying the Born-Markov approximation (note that the condition is automatically satisfied for the system discussed in this section), the new reduced density matrix $\tilde{\rho}_s(\tau)$ satisfies the Lindblad master equation
\begin{align}\label{IX81}
	\partial_\tau^{}\tilde{\rho}_{s}(\tau)=-i\left[H_{\rm sys}+H_{\rm d}(t),\tilde{\rho}_s(\tau)\right]+\frac{\kappa_l+\kappa_r}{2}\left[2 a\tilde{\rho}_s(\tau)a^\dagger -a^\dagger a\tilde{\rho}_s(\tau)-\tilde{\rho}_s(\tau)a^\dagger a\right].
\end{align}
Back to Eq.~(\ref{IX76}), the non-normalized $n$th-order equal-time correlation functions can be written as
\begin{align}\label{IX82}
	G_{rr}^{(n)}(0)=\Tr_{\rm S}\left[ b_r^{\dagger n} b_r^n\ \mathcal{U}(t)\rho_s(\tau)\mathcal{U}^\dagger(t)\right]=\Tr_{\rm S}\left[ b_r^{\dagger n} b_r^n\ \mathcal{U}(t)\tilde{\rho}_s(\tau)\mathcal{U}^\dagger(t)\right]=\Tr_{\rm S}\left[ b_r^{\dagger n} b_r^n  \mathring{\rho}_s(\tau)\right],
\end{align}
where $b_r\equiv \tilde{\eta}-iA_r=\customsqrt{\kappa_r}(\Omega/\kappa_r-ia)$ and $\mathring{\rho}_s(\tau)=\mathcal{U}(t)\tilde{\rho}_s(\tau)\mathcal{U}^\dagger(t)$. Using (\ref{IX81}), $\tau=t-t_+$, and $\mathcal{U}(t)=\exp(i\omega_d\mathbb{N}t)$, the time derivative of the reduced density matrix $\mathring{\rho}_s(\tau)$ is computed as
\begin{align}
	\partial_\tau^{}\mathring{\rho}_s(\tau)&=[\partial_\tau^{}\mathcal{U}(t)]\tilde{\rho}_s(\tau)\mathcal{U}^\dagger(t) + \mathcal{U}(t)[\partial_\tau^{}\tilde{\rho}_s(\tau)]\mathcal{U}^\dagger(t)+ \mathcal{U}(t)\tilde{\rho}_s(\tau)[\partial_\tau^{}\mathcal{U}^\dagger(t)]\nonumber\\
	&=-i[\mathcal{U}(t)H_{\rm sys}\mathcal{U}^\dagger(t)+\mathcal{U}(t)H_{\rm d}(t)\mathcal{U}^\dagger(t)-\omega_d\mathbb{N}, \mathring{\rho}_s(\tau)]+\frac{\kappa_l+\kappa_r}{2}\left[2 a\mathring{\rho}_s(\tau)a^\dagger -a^\dagger a\mathring{\rho}_s(\tau)-\mathring{\rho}_s(\tau)a^\dagger a\right]\nonumber\\
	&=-i[H_{\rm sys}-\omega_d\mathbb{N}+H_{\rm d}(0), \mathring{\rho}_s(\tau)]+\kappa\left[2 a\mathring{\rho}_s(\tau)a^\dagger -a^\dagger a\mathring{\rho}_s(\tau)-\mathring{\rho}_s(\tau)a^\dagger a\right]/2\equiv\mathcal{L}\mathring{\rho}_{s}(\tau),\label{IX83}
\end{align}
where $\mathcal{L}$ represents the Liouvillian operator and $\kappa=\kappa_l+\kappa_r$. In the last step, since the system Hamiltonian $H_{\rm sys}$ preserves $U(1)$ symmetry, $\mathbb{N}$ could be the total excitation number operator, i.e., $\mathbb{N}=a^\dagger a^{} + \sigma^\dagger \sigma^{}$, which satisfies $[H_{\rm sys}, \mathbb{N}]=0$ and $\mathcal{U}^\dagger(t)a\mathcal{U}^{}(t)=ae^{i\omega_d t}$. Due to $\tau=t-t_+=t+\infty\to+\infty$, Eq.~(\ref{IX82}) reads as
\begin{align}
	G_{rr}^{(n)}(0)=\langle \psi_{\text{out}}| b_r^{\dagger n}(t)b_r^{n}(t) |\psi_{\text{out}}\rangle=\Tr_{\rm S}\left[ b_r^{\dagger n} b_r^n  \mathring{\rho}_{s}(\tau)\right]=\Tr_{\rm S}\left[ b_r^{\dagger n} b_r^n  \mathring{\rho}_{\rm ss}\right],
\end{align} 
where $\mathring{\rho}_{\rm ss}=\lim\limits_{\tau\to+\infty}\mathring{\rho}_{s}(\tau)=\lim\limits_{\tau\to+\infty}\exp(\mathcal{L}\tau)\mathring{\rho}_s(0)=\lim\limits_{\tau\to+\infty}\exp(\mathcal{L}\tau)|0\rangle_s\langle0|$ representing the steady state of the master equation (\ref{IX83}). Note that for multi-cavity case, the corresponding master equation is given by Eq.~(\ref{eq49}). Accordingly, the normalized $n$th-order equal-time correlation functions are given by
\begin{align}\label{IX85}
	g_{rr}^{(n)}(0)=\frac{G_{rr}^{(n)}(0)}{\ \ [G_{rr}^{(1)}(0)]^n}=\frac{\langle \psi_{\text{out}}| b_r^{\dagger n}(t)b_r^{n}(t) |\psi_{\text{out}}\rangle}{\ \ \langle \psi_{\text{out}}| b_r^{\dagger }(t)b_r(t) |\psi_{\text{out}}\rangle^n}=\frac{\Tr_{\rm S}\left[ b_r^{\dagger n} b_r^n  \mathring{\rho}_{\rm ss}\right]}{\ \ \Tr_{\rm S}[ b_r^{\dagger } b_r  \mathring{\rho}_{\rm ss}]^n}.
\end{align}
Moreover, we note that $|\psi_{\text{out}}\rangle=S|\psi_{\text{in}}\rangle=\mathcal{N}\sum_{k=0}^{\infty}\eta^k[Sb_{r}^{\dagger k}(\omega_d)/k!]|0\rangle_r|0\rangle_l|0\rangle_s=\mathcal{N}\sum_{k=0}^{\infty}\eta^k|\tilde{\Psi}_{\rm out}^{(k)}\rangle$ and $\ln\mathcal{N}\propto\abs{\eta}^2$. To validate the Eq.~(3) of the main text obtained from the scattering matrix method under the weak coherent driving approximation (i.e., $\abs{\eta}\approx0$), we plug $|\psi_{\text{out}}\rangle$ into Eq.~(\ref{IX85}), taking $n=2$ as an example, which results in 
\begin{align}
	g_{rr}^{(2)}(0)&=\frac{\Tr_{\rm S}\left[ b_r^{\dagger n} b_r^n  \mathring{\rho}_{\rm ss}\right]}{\ \ \Tr_{\rm S}[ b_r^{\dagger } b_r  \mathring{\rho}_{\rm ss}]^n}=\frac{\langle \psi_{\text{out}}| b_r^{\dagger n}(t)b_r^{n}(t) |\psi_{\text{out}}\rangle}{\ \ \langle \psi_{\text{out}}| b_r^{\dagger }(t)b_r(t) |\psi_{\text{out}}\rangle^n}=\frac{\sum_{k,\ell=0}^{\infty}\abs{\eta}^{2k}\langle\tilde{\Psi}_{\rm out}^{(k)}|b_{r}^{\dagger 2}(t) b_{r}^{2}(t)|\tilde{\Psi}_{\rm out}^{(\ell)}\rangle}{[\sum_{k,\ell=0}^{\infty}\abs{\eta}^{2k}\langle\tilde{\Psi}_{\rm out}^{(k)}|b_{r}^\dagger(t) b_{r}(t)|\tilde{\Psi}_{\rm out}^{(\ell)}\rangle]^2\mathcal{N}^2}\nonumber\\
	&=\frac{\sum_{\ell=2}^{\infty}\abs{\eta}^{2\ell}\langle\tilde{\Psi}_{\rm out}^{(\ell)}|b_{r}^{\dagger 2}(t) b_{r}^{2}(t)|\tilde{\Psi}_{\rm out}^{(\ell)}\rangle}{[\sum_{\ell=1}^{\infty}\abs{\eta}^{2\ell}\langle\tilde{\Psi}_{\rm out}^{(\ell)}|b_{r}^\dagger(t) b_{r}(t)|\tilde{\Psi}_{\rm out}^{(\ell)}\rangle]^2\mathcal{N}^2}=\frac{\langle\tilde{\Psi}_{\rm out}^{(2)}|b_{r}^{\dagger 2}(t) b_{r}^{2}(t)|\tilde{\Psi}_{\rm out}^{(2)}\rangle+\abs{\eta}^2\langle\tilde{\Psi}_{\rm out}^{(3)}|b_{r}^{\dagger 2}(t) b_{r}^{2}(t)|\tilde{\Psi}_{\rm out}^{(3)}\rangle+o(\abs{\eta}^4)}{[\langle\tilde{\Psi}_{\rm out}^{(1)}|b_{r}^\dagger(t) b_{r}(t)|\tilde{\Psi}_{\rm out}^{(1)}\rangle+\abs{\eta}^2\langle\tilde{\Psi}_{\rm out}^{(2)}|b_{r}^\dagger(t) b_{r}(t)|\tilde{\Psi}_{\rm out}^{(2)}\rangle+o(\abs{\eta}^4)]^2\mathcal{N}^2}\nonumber\\
	&\xlongequal{\abs{\eta}\approx0}\frac{\langle\tilde{\Psi}_{\rm out}^{(2)}|b_{r}^{\dagger 2}(t) b_{r}^{2}(t)|\tilde{\Psi}_{\rm out}^{(2)}\rangle}{\langle\tilde{\Psi}_{\rm out}^{(1)}|b_{r}^\dagger(t) b_{r}(t)|\tilde{\Psi}_{\rm out}^{(1)}\rangle^2}+\abs{\eta}^2\frac{\langle\tilde{\Psi}_{\rm out}^{(3)}|b_{r}^{\dagger 2}(t) b_{r}^2(t)|\tilde{\Psi}_{\rm out}^{(3)}\rangle}{\langle\tilde{\Psi}_{\rm out}^{(1)}|b_{r}^\dagger(t) b_{r}(t)|\tilde{\Psi}_{\rm out}^{(1)}\rangle^2}+o(\abs{\eta}^4)=|\mathbb{P}_2|^2/|\mathbb{P}_1|^4+\abs{\eta}^2\mathbb{C}^{}+o(\abs{\eta}^4).\label{IX86}
\end{align}
In the last step of (\ref{IX86}), by applying $\mathcal{N}^2=1+o(\abs{\eta}^2)$ and $\langle\tilde{\Psi}_{\rm out}^{(k)}|b_{r}^{\dagger k}(t) b_{r}^{k}(t)|\tilde{\Psi}_{\rm out}^{(k)}\rangle\sim \langle\tilde{\Psi}_{\rm out}^{(k+1)}|b_{r}^{\dagger k}(t) b_{r}^k(t)|\tilde{\Psi}_{\rm out}^{(k+1)}\rangle$ (implying $\mathbb{C}\sim |\mathbb{P}_2|^2/|\mathbb{P}_1|^4$), we can obtain the coefficients for the zeroth-order and second-order terms in $\abs{\eta}$, while higher-order terms are not considered here. Clearly, the zeroth-order term, $|\mathbb{P}_2|^2/|\mathbb{P}_1|^4$, provides the analytical expression for Eq.~(3) in the main text.

\begin{figure}
	\includegraphics[width=17cm]{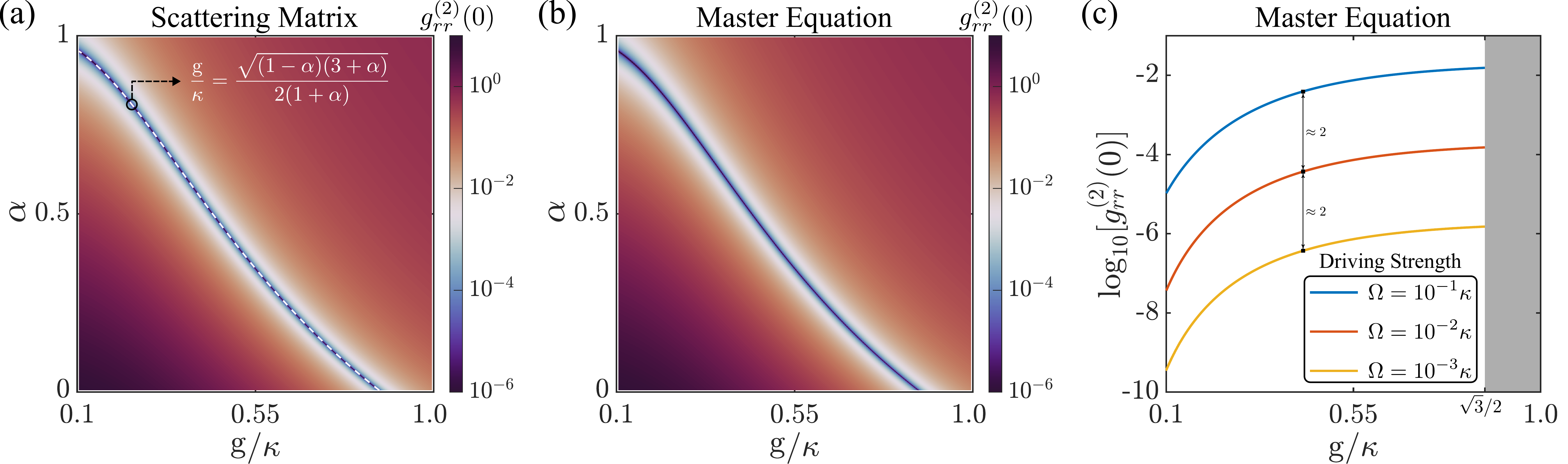}\\
	\caption{Panels (a) and (b) represent equal-time two-photon correlation $g_{rr}^{(2)}(0)$ versus the chirality $\alpha$ and the coupling strength $\text{g}/\kappa$, respectively calculated by the scattering matrix and the master equation with $\Omega=10^{-3}\kappa$. The dashed white line in (a) denotes the optimal parameter curve corresponding to $g_{rr}^{(2)}(0)=0$, obtained from the scattering matrix. Panel (c) exhibits $g_{rr}^{(2)}(0)$ as a function of the coupling strength for different driving strength, calculated by the master equation. In (c), all $\{\alpha,\text{g}/\kappa\}$, except those in the gray region, satisfy the optimal parameter curve. For all plots, the other parameters are the same as Fig.2(c) of the main text.
	}\label{Fig_Sup}
\end{figure}

In conclusion, from an analytical perspective, according to Eq.~(\ref{IX86}), we see that in the limit of $\abs{\eta}\ ({\rm or}\ \Omega)\to0$, the second-order equal-time correlation function obtained from the master equation is consistent with that from the scattering matrix method, i.e., 
\begin{align}
	\lim\limits_{\abs{\eta}\to0}\frac{\Tr_{\rm S}\left[ b_r^{\dagger n} b_r^n  \mathring{\rho}_{\rm ss}\right]}{\ \ \Tr_{\rm S}[ b_r^{\dagger } b_r  \mathring{\rho}_{\rm ss}]^n}=\lim\limits_{\abs{\eta}\to0}g_{rr}^{(2)}(0)=\frac{\abs{\mathbb{P}_2}^2}{\abs{\mathbb{P}_1}^4}.
\end{align}
From a numerical perspective, we will select a sufficiently small but non-zero driving strength $\Omega$, utilizing the open-source Python library QuTiP\,\cite{johansson_qutip_2012,johansson_qutip_2013} to numerically solve the steady state of the master equation (\ref{IX83}), and subsequently compute the corresponding correlation function. Figures \ref{Fig_Sup}(a) and \ref{Fig_Sup}(b) shows the comparison between master-equation based simulations with the scattering matrix method, and we see that in the weak coherent driving approximation, the QuTiP simulations agree perfectly with the scattering matrix based simulations. Note that in Figs.\,\ref{Fig_Sup}(a) and \ref{Fig_Sup}(b), to facilitate the comparison between the results obtained from the scattering matrix method and the master equation, we set $10^{-6}$ as the lower bound for the second-order equal-time correlation function, meaning any correlation function values smaller than the lower bound are set to $10^{-6}$. Additionally, to verify the form of the second-order small term in Eq.~(\ref{IX86}), i.e., $\abs{\eta}^2\mathbb{C}$, in Fig.\,\ref{Fig_Sup}(c), we select the system parameters $\{\alpha, \text{g}\}$ along the optimal parameter curve (where $|\mathbb{P}_2|^2/|\mathbb{P}_1|^4=0$) indicated by the white dashed line in Fig.\,\ref{Fig_Sup}(a). With the above parameters, according to Eq.~(\ref{IX86}), we have
\begin{align}\label{IX88}
	g_{rr}^{(2)}(0; \eta)\xlongequal{\abs{\eta}\approx0}|\mathbb{P}_2|^2/|\mathbb{P}_1|^4+\abs{\eta}^2\mathbb{C}^{}+o(\abs{\eta}^4)=\abs{\eta}^2\mathbb{C}^{}+o(\abs{\eta}^4)\approx\abs{\eta}^2\mathbb{C}^{},
\end{align}
which indicates that $\log_{10}[g_{rr}^{(2)}(0; \eta_1)]-\log_{10}[g_{rr}^{(2)}(0;\eta_2)]\approx 2\log_{10}\abs{\eta_1/\eta_2}$. Therefore, as shown by the spacing ($\approx 2$) between the black squares in Fig.\,\ref{Fig_Sup}(c), and given that $\Omega=\eta\sqrt{\kappa_r/2\pi}$, this numerically confirms the validity of Eq.~(\ref{IX88}). Specifically, the lowest-order perturbation in the correlation function, stemming from the sufficiently small but non-zero driving strength, is proportional to $|\Omega|^2$ ($\propto |\eta|^2$). 

As a result, whether from an analytical or numerical perspectives, the scattering matrix method is fully valid and practical under the weak driving approximation.

\vspace{-10pt}
\section{Matrix Product States Simulations for the Multi-Cavity Case}\label{X}
\begin{figure}
	\includegraphics[width=17cm]{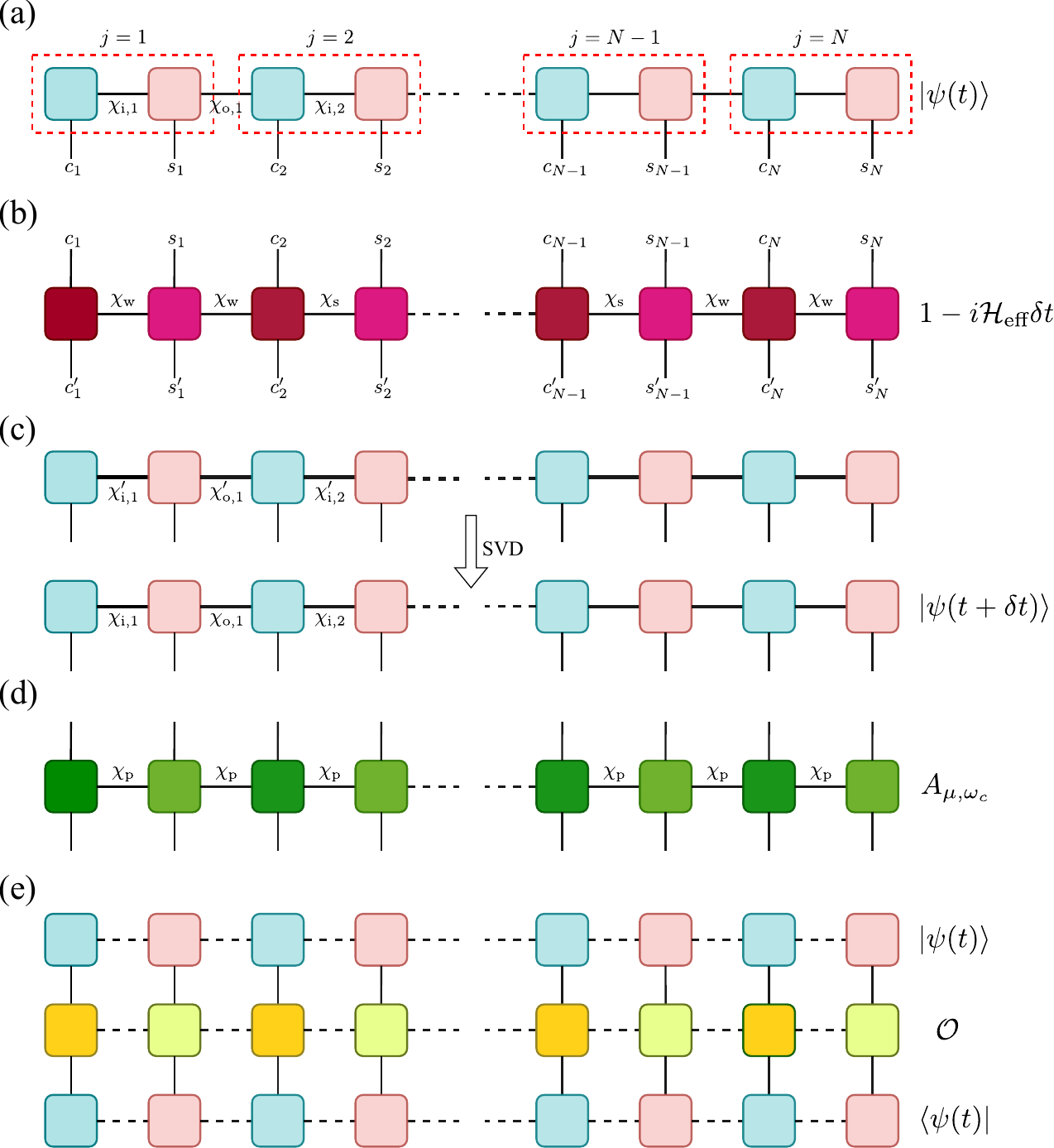}\\
	\caption{(a) In MPS representation, the state $|\psi(t)\rangle$ at time $t$ is presented pictorially as a tensor network, where the cyan and pink squares symbolize the sets of local matrices $B^{[j]c_j}$ ($B^{[j]s_j}$) associated with the $j$-th local system surrounded by the corresponding red dashed lines. The connecting lines or bonds between the squares represent the contraction of these local tensors, yielding the state $|\psi(t)\rangle$, and the bonds have dimensions $\chi_{\text{i},j}$ and $\chi_{\text{o},j}$. The open-ended lines on the cyan squares and pink squares correspond to the Hilbert space of the cavity $c_j$ and the atom $s_j$, respectively. (b) The MPO representation of $1-i\mathcal{H}_{\text{eff}}\delta t$ with two types of bond dimensions: $\chi_\text{w}$ and $\chi_\text{s}$. (c) After the application of the time evolution (b) or jump (d) MPOs the resulting MPS has larger bond dimensions, e.g., $\chi_{\text{i},1}^\prime=\chi_\text{w}\times \chi_{\text{i},1}$ and $\chi_{\text{i},2}^\prime=\chi_\text{s}\times \chi_{\text{i},2}$, and is subsequently truncated to the initial bond dimension using the SVD. (d) The MPO representation of a quantum jump operator $A_{\mu,\omega_c}$ with the bond dimension $\chi_\text{p}$. (e) Measurement of observables. At time $t$, we may evaluate an observable $\mathcal{O}$ by sandwiching the corresponding MPO between the MPSs representing $|\psi(t)\rangle$ and $\langle\psi(t)|$, so that the corresponding tensor contraction yields the expectation value $\langle\psi(t)|\mathcal{O}|\psi(t)\rangle$. For the operators, $1-i\mathcal{H}_\text{eff}\delta t$, $A_{\mu,\omega_c}$, and $\mathcal{O}$, the corresponding MPOs have different colors respectively, as shown in (b), (c), and (d). In each MPO representation, the corresponding squares are color-coded using two distinct colors to differentiate between the physical indices of the cavity and atom modes. 
	}\label{fig3}
\end{figure}
With the increasing number of cavities, the Hilbert space dimension of the total system grows exponentially, making traditional numerical methods based on the tools like QuTiP impractical. In this section, we address this challenge by employing matrix product states (MPS) simulations. By comparing the results obtained from the MPS simulations with the second-order correlation function derived from the $S$-matrix method, we aim to validate the consistency and accuracy of the $S$-matrix method even for large-scale systems. In our system, the MPS simulations are effective for the two main reasons below:
\begin{itemize}
	\item The weak driving approximation employed in this study ensures lower photon numbers within the total system. As a consequence, the Hilbert space of the cavity mode can be truncated to a finite number of photons.
	\item For a general driven-dissipation systems, it should reach a steady state over time and there will not be an indefinite growth of entanglement\,\cite{manzoni_simulating_2017}. 
\end{itemize}

In particular, the wave function of many-body system can be brought into an MPS form, as illustrated in Fig.\,\ref{fig3}(a), i.e.,
\begin{align}\label{eq48}
	|\psi\rangle_{\text{MPS}}=\sum_{\{c_1,s_1\},\ldots,\{c_N,s_N\}}B^{[1]c_1}B^{[1]s_1}B^{[2]c_2}B^{[2]s_2}\cdots B^{[N]c_N}B^{[N]s_N}|c_1,s_1,c_2,s_2,\ldots,c_N,s_N\rangle,
\end{align}
where $c_j$ and $s_j$ represent the physical indices of the $j$-th cavity mode and the $j$-th two-level atom mode, respectively. Here, the superscript index in square brackets $[j]$ denotes the $j$-th local system that contains a cavity and a two-level atom, and $B^{[j]c_j}\in\mathbb{C}^{\chi_{\text{o},j-1}\times\chi_{\text{i},j}}$ ($B^{[j]s_j}\in\mathbb{C}^{\chi_{\text{i},j}\times\chi_{\text{i},j}}$) are the matrices corresponding to the $j$-th system and to the local state $|c_j\rangle$ ($|s_j\rangle$).  Note that $B^{[1]c_1}$ ($B^{[N]s_N}$) are row (column) vectors of size $\chi_{\text{i},1}$ ($\chi_{\text{i},N}$) to make the wavefunction coefficients scalar and so $\chi_{\text{o},0}=\chi_{\text{o},N}=1$. The required bond dimension $\chi_{\cdot,\cdot}$ depends on the amount of bipartite entanglement included in $|\psi_{\text{MPS}}\rangle$. We also define $\chi=\max_j\{\chi_{\text{i},j},\chi_{\text{o},j}\}$ as the maximum bond dimension of the state $|\psi_{\text{MPS}}\rangle$. Besides, the Hilbert space of the cavity mode is truncated to accommodate maximum number of photons $N_{\text{max}}^{\text{cav}}=5$, while the atomic local Hilbert space dimension is 2. 

In our MPS treatment of the cavity-atom array model, we utilize a quantum trajectory method to accurately model of the long time dynamics of the master equation, and the master equation for our system can be written as\,\cite{PhysRevA.91.042116}
\begin{align}\label{eq49}
	\dot{\rho}=-i\left(\mathcal{H}_{\text{eff}}\rho-\rho \mathcal{H}_{\text{eff}}^\dagger\right)+\sum_{\mu=l,r}A_{\mu,\omega_c}^{}\rho A_{\mu,\omega_c}^\dagger,
\end{align}
where
\begin{align}\label{eq50}
	\mathcal{H}_{\text{eff}}=\sum_{j=1}^N{[\Delta(a_j^\dagger a_j^{}+\sigma_j^\dagger\sigma_j^{})+\text{g}(a_j^\dagger\sigma_j^{}+\sigma_j^\dagger a_j^{})]}+\mathcal{H}_{\text{drive}}-i\sum_{j=1}^N\frac{\kappa_l+\kappa_r}{2}a_{j}^\dagger a_{j}^{}\!-\!i\sum_{j>k}^Ne^{i\omega_c(x_j-x_k)/v_r}[\kappa_ra_{j}^\dagger a_{k}^{}+\kappa_la_{k}^\dagger a_{j}^{}],
\end{align}
and $A_{\mu,\omega_c}$ also are the “jump” operators associated with the dissipation of the cavities resulting from emission into the waveguide. Note that $\mathcal{H}_{\text{drive}}=\Omega\sum_{j=1}^N{(e^{i\omega_cx_j/v_r}a^\dagger_j+e^{-i\omega_cx_j/v_r}a_j)}$ corresponds to the weak coherent state input, and we chose the drive amplitude $\Omega=5\times10^{-3}\kappa$ in all of the other MPS results presented in this work. In the quantum trajectory method, the wave function is divided into deterministic evolution under $\mathcal{H}_{\text{eff}}$ and stochastic quantum jumps made by applying jump operators $A_{\mu,\omega_c}$. 

For the deterministic evolution, the corresponding evolution operator is $\exp(-i\mathcal{H}_{\text{eff}}\delta t)$. Here, we take $\delta t \ll 1/\kappa$ and use the first order approximation $\exp(-iH_{\text{eff}}\delta t)\approx 1-i\mathcal{H}_{\text{eff}}\delta t$. As a consequence, the evolution operator has a compact matrix product operators (MPO) form with $\chi_{\text{w}}=4$ and $\chi_{\text{s}}=6$, as illustrated in Fig.\,\ref{fig3}(b). Subsequently, we can write $\mathcal{H}_{\text{eff}}=W^{[1]c}W^{[1]s}\cdots W^{[N]c}W^{[N]s}$, where $W^{[j]c}=\sum_{c_j^\prime,c_j^{}}W^{c_j^\prime,c_j^{}}|c_j^\prime\rangle\langle c_j|$ and $W^{[j]s}=\sum_{s_j^\prime,s_j^{}}W^{s_j^\prime,s_j^{}}|s_j^\prime\rangle\langle s_j|$. Here, the matrices $W^{[j]c}$ and $W^{[j]s}$ for $1<j<N$ are given by 
\begin{align}\label{eq51}
	W^{[1<j<N]c}=\mqty[\mathcal{I}_j^{c}& -i\kappa_l e^{i\phi_j} a_j^\dagger&-i\kappa_re^{i\phi_j} a_j&\mathcal{H}_{\text{loc},j}^{c}&a_j&a_j^\dagger\\
	0&e^{i\phi_j}\mathcal{I}_j^c&0&a_j&0&0\\
	0&0&e^{i\phi_j}\mathcal{I}_j^c&a_j^\dagger&0&0\\
	0&0&0&\mathcal{I}_j^c&0&0
	],\ 
	W^{[1<j<N]s}=\mqty[\mathcal{I}_j^{s}& 0&0&0&0&0\\
	0&\mathcal{I}_j^s&0&0&0&0\\
	0&0&\mathcal{I}_j^s&0&0&0\\
	\mathcal{H}_{\text{loc},j}^s&0&0&\mathcal{I}_j^s&\text{g}\sigma_j^\dagger&\text{g}\sigma_j
	]^T.
\end{align}
The matrices $W^{[j]c}$ and $W^{[j]s}$ for $j=1$ are given by 
\begin{align}\label{eq52}
	W^{[1]c}=\mqty[\mathcal{I}_1^{c}& -i\kappa_le^{i\phi_1} a_1^\dagger&-i\kappa_re^{i\phi_1} a_1&\mathcal{H}_{\text{loc},1}^{c}
	],\ 
	W^{[1]s}=\mqty[\mathcal{I}_j^{s}& 0&0&\mathcal{H}_{\text{loc},1}^s\\
	0&\mathcal{I}_j^s&0&\frac{i\text{g}e^{-i\phi_1}}{\kappa_l}\sigma_1\\
	0&0&\mathcal{I}_j^s&\frac{i\text{g}e^{-i\phi_1}}{\kappa_r}\sigma_1^\dagger\\
	0&0&0&\mathcal{I}_j^s
	].
\end{align}
The matrices $W^{[j]c}$ and $W^{[j]s}$ for $j=N$ are given by 
\begin{align}\label{eq53}
	W^{[N]c}=\mqty[\mathcal{I}_j^{c}& \text{g}a_N^\dagger&\text{g} a_N^{}&\mathcal{H}_{\text{loc},N}^{c}\\
	0&0&0&a_N\\
	0&0&0&a_N^\dagger\\
	0&0&0&\mathcal{I}_j^c
	],\ 
	W^{[N]s}=\mqty[\mathcal{H}_{\text{loc},N}^s& \sigma_N^{}&\sigma_N^\dagger&\mathcal{I}_N^s
	]^T,
\end{align}
where $\phi_j=\omega_c(x_{j+1}-x_j)/v_r$, $\mathcal{I}_{j}^{c}$ ($\mathcal{I}_{j}^{s}$) are the identity operators of the $j$-th cavity (atom) mode, and the $\mathcal{H}_{\text{loc},j}^c$ ($\mathcal{H}_{\text{loc},j}^s$) contain all the local terms about the $j$-th cavity (atom) mode in $\mathcal{H}_\text{eff}$, i.e., $\mathcal{H}^c_{\text{loc},j}=(\Delta-i\kappa/2)a_j^\dagger a_j^{}+\mathcal{H}_{\text{drive}}^{(j)}$ ($\mathcal{H}^s_{\text{loc},j}=\Delta\sigma_j^\dagger\sigma_j^{}$). Meanwhile, it is enough to replace $W^{[1]c}$ with
\begin{align}\label{eq54}
	W^{[1]c}_{\text{t.e.}}=\mqty[-i\delta t\mathcal{I}_1^{c}& -\delta t\kappa_le^{i\phi_1} a_1^\dagger&-\delta t\kappa_re^{i\phi_1} a_1&\mathcal{I}_j^c-i\delta tH_{\text{loc},1}^{c}
	]
\end{align}
to acquire the desired MPO, i.e., $1-i\mathcal{H}_{\text{eff}}\delta t=W_{\text{t.e.}}^{[1]c}W^{[1]s}\cdots W^{[N]c}W^{[N]s}$, corresponding to Fig.\,\ref{fig3}(b). Given a pure state $|\psi(t)\rangle$ at time $t$, we obtain a new state at time $t+\delta t$ by applying the MPO representation of the evolution operator to the MPS representation of $|\psi(t)\rangle$ via a tensor contraction over the physical indices $c_j$ and $s_j$ of the MPS and MPO, i.e., $|\psi^{(1)}(t+\delta t)\rangle=(1-i\mathcal{H}_{\text{eff}}\delta t)|\psi(t)\rangle$. This generates a new MPS with higher bond dimension, as the bond dimension of the MPO, $\chi_\text{w}$ or $\chi_{\text{s}}$, multiples the bond dimension of the original MPS, such as $\chi_{\text{i},j}$ and $\chi_{\text{o},j}$. However, for the calculation, we need to use the singular value decompositions (SVD) to find a low rank approximation of the matrices $B^{[j]c_j}$ and $B^{[j]s_j}$ in the MPS representation, as shown in Fig.\,\ref{fig3}(c), and it does work due to the limited entanglement. 

\begin{figure}
	\includegraphics[width=17cm]{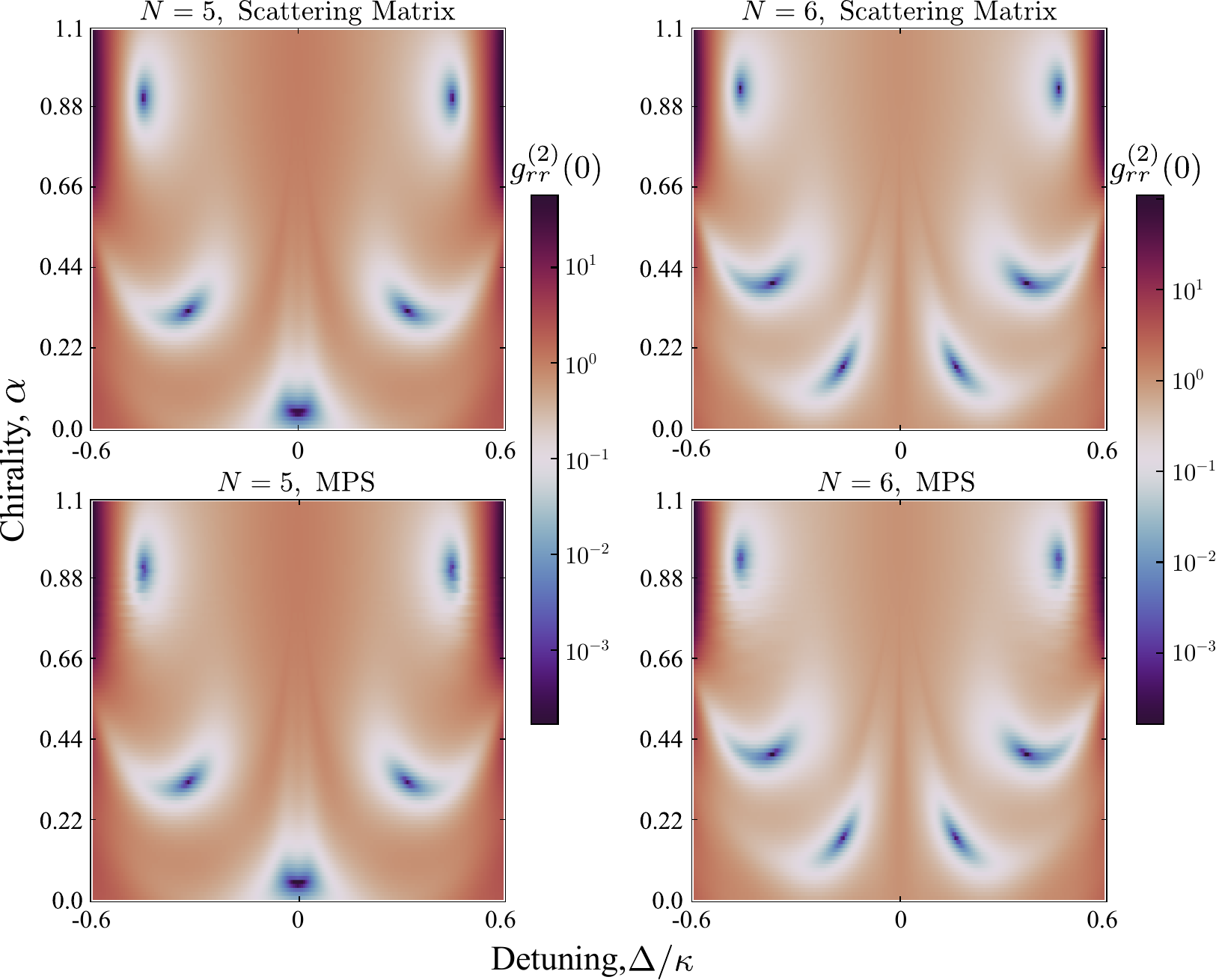}\\
	\caption{MPS verification of the $S$-matrix result for the number of cavities $N=5$ and $N=6$. The color of the heat map corresponds to $g_{rr}^{(2)}(0)$. All parameters are the same as in Fig.\,3 of the main text.
	}\label{fig4}
\end{figure}

Then, according to quantum trajectory method\,\cite{qt}, the state $|\psi(t+\delta t)\rangle$ at time $t+\delta t$ for a single trajectory is either kept and renormalized, or a jump is applied. This means that
\begin{itemize}
	\item With probability $\delta p=\langle\psi^{(1)}(t+\delta t)|\psi^{(1)}(t+\delta t)\rangle$
		\begin{align}\label{eq55}
			|\psi(t+\delta t)\rangle=\frac{|\psi^{(1)}(t+\delta t)\rangle}{\sqrt{\langle\psi^{(1)}(t+\delta t)|\psi^{(1)}(t+\delta t)\rangle}}.
		\end{align}
	\item With probability $1-\delta p$
		\begin{align}\label{eq56}
			|\psi(t+\delta t)\rangle=\frac{A_{\mu,\omega_c}|\psi(t)\rangle}{\sqrt{\langle\psi(t)|A_{\mu,\omega_c}^\dagger A^{}_{\mu,\omega_c}|\psi(t)\rangle}},\quad
		\end{align}
where we select one \textit{particular} $\mu$, which is taken from all of the possible $\mu$ with probability 
		\begin{align}\label{eq57}
			\Pi_\mu=\frac{\langle\psi(t)|A_{\mu,\omega_c}^\dagger A^{}_{\mu,\omega_c}|\psi(t)\rangle}{\sum_{\mu=l,r}\langle \psi(t)|A_{\mu,\omega_c}^\dagger A^{}_{\mu,\omega_c}|\psi(t)\rangle}.
		\end{align}
\end{itemize} 
Similarly, the jump operators $A_{\mu,\omega_c}$ also can be written in compact MPO form with bond dimension $\chi_{\text{p}}=2$, as shown in Fig.\,\ref{fig3}(d), and we can write $A_{\mu,\omega_c}=Z_{\mu}^{[1]c}Z^{[1]s}_{}Z_{\mu}^{[2]c}Z^{[2]s}_{}\cdots Z_{\mu}^{[N]c}Z^{[N]s}_{}$ for $\mu\in\{l,r\}$ with
\begin{align}\label{eq58}
	Z_{l}^{[1<j\le N]c}=\mqty[\mathcal{I}_j^c & \sqrt{\kappa_l}e^{+i\omega_cx_j/v_r}a_j\\
						0&\mathcal{I}_j^c],\ Z_{r}^{[1<j\le N]c}=\mqty[\mathcal{I}_j^c & \sqrt{\kappa_r}e^{-i\omega_cx_j/v_r}a_j\\
						0&\mathcal{I}_j^c],\ Z^{[1\le j< N]s}=\mqty[\mathcal{I}_j^s & 0\\
						0&\mathcal{I}_j^s],
\end{align}
and end vectors 
\begin{align}\label{eq59}
	Z_{l}^{[1]c}=\mqty[\mathcal{I}_1^c&\sqrt{\kappa_l}e^{+i\omega_cx_1/v_r}a_1 ],\ Z_{r}^{[1]c}=\mqty[\mathcal{I}_1^c&\sqrt{\kappa_r}e^{-i\omega_cx_1/v_r}a_1 ],\ Z^{[N]s}=\mqty[0&\mathcal{I}_N^s]^T.
\end{align}
Note that, Eq.\,(\ref{eq56}) could be denoted as the operation that an MPO is applied to an MPS and then performs a state compression. In terms of a single trajectory, we can repeat these steps of Eqs.\,(\ref{eq55}-\ref{eq57}) to obtain a new state at time $t=T_{\text{ss}}$, where $T_{\text{ss}}$ represents an evolving time from the vacuum to the steady state. Finally, in order to compute a particular physical quantity at time $t$, such as $\langle \mathcal{O}\rangle_t=\Tr[\mathcal{O}\rho(t)]$, we need to do a statistical average to the expectation value $\langle\psi(t)|\mathcal{O}|\psi(t)\rangle$ over all of the trajectories, i.e.,
\begin{align}\label{eq60}
	\langle\mathcal{O}\rangle_t\approx\overline{\langle\psi(t)|\mathcal{O}|\psi(t)\rangle},
\end{align}
where the symbol $\overline{X}$ represents the statistical average over all trajectories of $X$. Here, the expectation value is expressed as the operation that the corresponding MPO is sandwiched between $|\psi(t)\rangle$ at the top and $\langle\psi(t)|$ at the bottom, as shown in Fig.\,\ref{fig3}(e). To this end, the second-order correlation function can be written as
\begin{align}\label{eq61}
	g_{rr}^{(2)}(0)=\frac{\langle b_r^\dagger b_r^\dagger b^{}_r b^{}_r\rangle_{t=T_{\text{ss}}}}{\langle b_r^\dagger b_r^{}\rangle_{t=T_{\text{ss}}}\times\langle b_r^\dagger b_r^{}\rangle_{t=T_{\text{ss}}}}\approx\frac{\overline{\langle\psi(T_{\text{ss}})|b_r^\dagger b_r^\dagger b_r b_r|\psi(T_{\text{ss}})\rangle}}{\overline{\langle\psi(T_{\text{ss}})|b_r^\dagger b_r|\psi(T_{\text{ss}})\rangle}\times\overline{\langle\psi(T_{\text{ss}})|b_r^\dagger b_r|\psi(T_{\text{ss}})\rangle}},
\end{align}
where $b_r=\Omega/\customsqrt{\kappa_r}-iA_{r,\omega_c}$. Similarly, according to Eq.\,(\ref{eq54}), we find that $b_r$ also can be written in compact MPO form. In the MPS simulations, we used $T_{\text{ss}}=10^3/\kappa$, the number of trajectories $N_{\text{traj}}=10^2$, and $\chi=50$.

Finally, as shown in Fig.\,\ref{fig4}, we see that the MPS simulations agree perfectly with the scattering matrix method for different $N$, thereby validating the scattering matrix method.

\end{document}